\documentclass[article]{achemso}[article]
\SectionNumbersOn
\RequirePackage{etex} 
\usepackage{amsmath}
\usepackage{amssymb}
\usepackage{sansmath}
\usepackage{graphicx} 
\usepackage{ragged2e}

\usepackage{mathtools}

\usepackage[numbers,sort&compress]{natbib}
\bibpunct[, ]{[}{]}{,}{n}{,}{,}
\usepackage{xcolor}
\usepackage[british]{babel}
\usepackage{chemgreek}
\usepackage[version=3,version=4]{mhchem} 
\usepackage[modules=all]{chemmacros}

\usepackage{epstopdf}
\usepackage[caption=false]{subfig}
\usepackage{bm}

\usepackage{tikz}
\usetikzlibrary{arrows}
\usetikzlibrary {graphs,shapes.geometric,perspective,arrows.meta}
\usepackage{tikz-network}
\usepackage{braket}

\usepackage{blindtext}
\usepackage{hyperref}
\usepackage[export]{adjustbox}

\author{Nicolás Salas}
\affiliation[Icesi]
{Department of Pharmaceutical and Chemical Sciences, Universidad Icesi, Cali, Colombia}
\author{Justin López}
\affiliation{Department of Pharmaceutical and Chemical Sciences, Universidad Icesi, Cali, Colombia}
\author{Carlos A. Arango}
\affiliation{Department of Pharmaceutical and Chemical Sciences, Universidad Icesi, Cali, Colombia}
\email{caarango@icesi.edu.co}

\title{The graph automorphism group of the dissociation microequilibrium of polyprotic acids}

\begin{document}

\maketitle

\begin{abstract}

The dissociation micro-states (DMSs) of an $N$-protic acid are described using set theory notation, which facilitates the mathematical description of the dissociation micro-equilibrium (DME). In particular, the DME constants are easily obtained in terms of the dissociation equilibrium constants and the molar fractions of the DMSs. Representing of the DMEs in terms of graph theory allows to identify permutations between DMSs that preserve the vertex-edge connectivity of the graph. These permutations, along with their compositions, led to the identification of the direct product $C_2\times S_N$ of the cyclic group $C_2$, and the symmetric group $S_N$, as the graph automorphism group of the microdissociation of $N$-protic acids for $N=1,2,\dots,6$. In this context the microdissociations are associated with the $C_2$ group while the tautomerizations are related to the $S_N$ group. 

\end{abstract}

\section{Introduction}

Chemical applications of graph theory have origin in the earlier works on  the enumeration of chemical isomers \cite{Cayley1874,Cayley1875, Brunel1894,Brunel1898,Rouvray1974,Rouvray1989,Gutman2014} and the depiction of molecules \cite{Sylvester1878,Rouvray1989}. To date, there are numerous chemical applications of graph theory in research fields such as biochemistry, chemical kinetics, catalysis, quantum chemistry, NMR spectroscopy, chemoinformatics, and new drug discovery, among others \cite{Balaban1976,Trinajstic1992,Bonchev1994a,Bonchev1994b,Hashemi2022}. In chemoinformatics, graph theory has been instrumental in finding similarities between molecules to discover new drugs \cite{Droschinsky2022,Branco2023}. Applications of graph theory to chemical reaction networks have allowed the description and representation of complex reaction mechanisms using topological and complexity indices \cite{Clarke1974,Clarke1980,Temkin1995,Temkin1996,Zeigarnik1996,Feinberg2019,Futamata2021}. Reaction networks and graph theory have been used to study the dissociation of weak acids and bases. For example, the method of exponential polytopes has been employed to obtain approximate formulas for the pH of monoprotic weak acids \cite{Clarke1980}. Graph kernels have been utilized to estimate acid/base dissociation constants in molecules of biopharmaceutical interest \cite{Rupp2010}. Graph convolutional neural networks have been used for predicting the p$K_\mathrm{a}$ values and protonation state distribution of molecules of pharmaceutical interest \cite{Johnston2023}. In homogeneous catalysis, graph-theoretical methodology has allowed scientist to study the chemical reaction space of multicomponent mixtures \cite{Hashemi2022}. In quantum chemistry, weighted-graph-theoretical approaches are used to compute contributions from many-body approximations in post-Hartree-Fock molecular dynamics \cite{Zhang2021}. 

In biochemistry, graph theory methods have been employed to analyze the covalent and non-covalent bond networks in proteins, allowing for the identification of flexible and rigid regions in these biomolecules \cite{Jacobs2001}. Additionally, in biochemistry, graph convolutional networks have been used for automated function prediction of proteins from the protein structure \cite{Gligorijevic2021}. Dynamical graph analysis of the hydrogen bond network has allowed to study structural changes in membrane proteins, and the identification of protein groups relevant for proton transfer activity \cite{Siemers2019}. In molecular biology, significant research has utilized graph theory and discrete mathematics to explore the relationship between sequence, structure, and function in biological molecules \cite{Jonoska2014}. Graph theory methods have been particularly effective in analyzing RNA secondary structures, providing insights into their functional roles and interactions \cite{Heitsch2014}. Furthermore, graph-theoretical complexes have been used to model DNA self-assembly of nanoscale geometric constructs, yielding promising results for future nanotechnological applications in drug delivery, biosensors, and biomolecular computing \cite{Ellis-Monaghan2014}. 

Algebraic graph theory applies algebraic methods to graphs \cite{Godsil2001}. The connection between graphs and group theory is one of the main branches of algebraic graph theory \cite{Biggs1993}. The permutations of the vertices of a graph that maintain the edge-vertex connectivity endowed with the composition operation define the automorphism group of the graph. In chemistry, obtaining the automorphism group of a an associated graph is important in spectroscopy, quantum chemistry, structural elucidation and prediction of NMR spectrum of molecules, and in the characterization of molecular complexity, among other applications \cite{Razinger1993, Balasubramanian1994,Balasubramanian1995, Randic2003,Tratch2008}.  

A polyprotic Br{\o}nsted-Lowry acid is a substance capable of donating more than one proton \cite{petrucci2023,atkins2008}. These protons are released one by one in the consecutive (macroscopic) model, from the most acidic to the least acidic. The concentration of the chemical species involved in the consecutive dissociation model are obtained from the macroscopic equilibrium constants, the auto-ionization of water, and the balances of mass and charge of the acid dissolution \cite{Harris2019, Skoog2021,Denbigh1981,Stumm2012}. In the non-consecutive (microscopic) dissociation model, the $N$ protons are released independently. The concentrations of the chemical species involved in the non-consecutive dissociation are obtained from the micro-equilibrium constants, the auto-ionization of water, and the balances of charge and mass \cite{Adams1916,Petrossyan2019}. Micro-equilibrium constants and concentrations are important in biochemistry and pharmaceutical sciences \cite{Yatsimirskii1978,Noszal1986,Hernandez1997,Mazak2016, Scholz2018,Palla2019,Li2023}. The relations between microscopic and macroscopic  equilibrium constants of polyprotic acids were developed by Hill \cite{Hill1943}. 

This work is organized as follows. Sections \ref{sec:consecutive_model} and \ref{sec:non_consecutive} are used to introduce the consecutive and non-consecutive acid dissociation models, respectively, along with the corresponding notation based on sets. Section \ref{sec:micro-mracro-equilibrium-theo} demonstrates that the use of a notation based on set theory simplifies the complicated expression obtained by Hill in terms of indices, allowing for a general expression to be derived for the micro-equilibrium constants. These are expressed in terms of the equilibrium constants and the molar fractions of the tautomers of the acid's dissociation states. Examples of the diprotic and triprotic acids are provided in 
Section \ref{sec:results_micro_macro}. In Section \ref{sec:results_graph_automorphism}, the use of graph theory to represent the microdissociation of polyprotic acids and to obtain their graph automorphism group is explored. The aqueous dissociation of $N$-protic acids, with $N=1,2,\dots,6$ is thoroughly examined, identifying $\mathrm{Aut}(G_N)=C_2\times S_N$, as the automorphism group of these $N$-protic acid dissociation.   

\section{Theory and Methods}

\subsection{Consecutive dissociation of polyprotic acids}\label{sec:consecutive_model}

The aqueous dissociation equilibrium of a polyprotic weak acid $\ce{H_{\textit{N}}B}$ is given by $N$ consecutive acid dissociations plus the water auto-ionization,  
\begin{align}
    \ce{Z_{\nu-1} + H2O &<--> H3O+ + Z_{\nu}}, \hspace{1cm} 1\leq \nu \leq N,\label{eq:HNB_rx}\\
    \ce{2 H2O &<--> H3O+ + OH-}\label{eq:H2O_rx},
\end{align}
respectively. The equilibria displayed by equations \eqref{eq:HNB_rx} and \eqref{eq:H2O_rx} are effective equilibria since the $N$ protons of $\ce{H_{\textit{N}}B}$ can dissociate separately and not necessarily consecutively \cite{Adams1916,Scholz2018,Petrossyan2019}. The acid $\ce{H_{\textit{N}}B}$ has $N+1$ deprotonation states (DS)
\begin{equation}\label{eq:protonation_states}
    \ce{Z}_\nu={\ce{H_{\textit{N-\nu}}B^{\nu-}}},\hspace{1cm}{\nu=0,1,2,\dots,N},
\end{equation}
with ${\ce{Z_0}}={\ce{H_{\textit{N}}B}}$ as the fully protonated state, and $\ce{Z}_N={\ce{B}^{\textit{N}-}}$ as the fully deprotonated state. The activities of the DSs $\ce{Z_\nu}$ are given by
$a_\nu(\ce{Z_\nu})=\gamma_\nu\ce{[Z_\nu]}/C^\circ$, with $\gamma_\nu$ as the molar activity coefficient, $\ce{[Z_\nu]}$ as the molar concentration, and $C^\circ=1\,\mathrm{M}$. The activities of the hydronium and hydroxide ions are given by $a(\ce{H3O+})=\gamma_{\ce{H3O+}}\ce{[H3O+]}/C^\circ$ and $a(\ce{OH-})=\gamma_{\ce{OH-}}\ce{[OH^-]}/C^\circ$, with $\ce{[H3O+]}$ and $\ce{[OH^-]}$ as the molar concentrations of the hydronium and hydroxide ions.

The equilibrium state of a $N$-protic acid at analytical concentration $C_\mathrm{a}$ is mathematically characterized by $N+3$ equations: the $N$ equilibrium constants, the water autoionization constant, and the balances of charge and mass 
\begin{align}
    K_\nu&=\frac{a(\ce{Z_\nu})a(\ce{H3O+})}{a(\ce{Z_{\nu-1}})}, \hspace{1cm}\nu=1,2,\dots,N,\label{eq:K}\\
    K_\mathrm{w}&=a(\ce{H3O+})a(\ce{OH-}),\label{eq:Kw_activities}\\
    \ce{[H3O+]}&=\ce{[OH^-]}+\sum_{\nu=0}^N{\nu\ce{[Z_\nu]}},\label{eq:charge_balance}\\
    C_\mathrm{a}&=\sum_{\nu=0}^N{\ce{[Z_\nu]}},\label{eq:mass_balance}    
\end{align}
respectively. In the case of diluted solutions, it is valid to use the approximations $a(\ce{Z_\nu})\approx\ce{[Z_\nu]}/C^\circ$, $a(\ce{H3O+})\approx\ce{[H3O+]}/C^\circ$, and $a(\ce{OH-})\approx\ce{[OH^-]}/C^\circ$. The use of the biochemical standard state,  $C^\standardstate=10^{-7}C^\circ$, to define the dimensionless quantities $z_\nu=\ce{[Z_\nu]}/C^\standardstate$, $x=\ce{[H3O+]}/C^\standardstate$, and $y=\ce{[OH^-]}/C^\standardstate$, allows to rewrite equations \eqref{eq:K} to \eqref{eq:mass_balance} as
\begin{align}
    k_\nu&=\frac{z_\nu x}
    {z_{\nu-1}}, \hspace{1cm}\nu=1,2,\dots,N,\label{eq:k2}\\
    1&=xy,\label{eq:kw}\\
    x&=y+\sum_{\nu=0}^N{\nu z_\nu},\label{eq;charge_balance2}\\
    c_\mathrm{a}&=\sum_{\nu=0}^N{z_\nu},\label{eq:mass_balance2}    
\end{align}
with $c_\mathrm{a}=C_\mathrm{a}/C^\standardstate$, $k_\mathrm{w}=1$, and $k_\nu=10^7 K_\nu$.

This $N$-protic acid has $N$ distinguishable sites, each capable of being either occupied by a proton or left unoccupied. In this context, the dissociation state $\mathrm{Z}_\nu$ is characterized by having $N-\nu$ occupied sites and $\nu$ unoccupied sites. The sites of $\ce{H_{\textit{N}}B}$ are in one-to-one correspondence with the elements of the set $\mathbb{S}_N=\{1,2,\dots,N\}$. 

There are two distinct deprotonation schemes: consecutive and non-consecutive. In consecutive or sequential acid dissociation, protons on the occupied sites are released in a specific, predetermined order. Formally, this order is given by an enumeration of $\mathbb{S}_N$, $\mathcal{E}=(\nu_1,\nu_2,\dots,\nu_N)$, where 
\begin{align}
    \nu_1&\in \mathbb{S}_N,\nonumber\\
    \nu_2&\in \mathbb{S}_N-\{\nu_1\},\nonumber\\
    \nu_3&\in \mathbb{S}_N-\{\nu_1,\nu_2\},\dots,\nonumber\\
    \nu_n&\in \mathbb{S}_N-\{\nu_1,\nu_2,\dots,\nu_{n-1}\},\dots,\nonumber\\
    \nu_N&\in \mathbb{S}_N-\{\nu_1,\nu_2,\dots,\nu_{N-1}\}.\nonumber
\end{align}
In the consecutive dissociation there is a one-to-one mapping between the states $\ce{Z}_\mu$ and the sets $\mathbb{S}_N-\{\nu_1,\nu_2,\dots,\nu_\mu\}$, with $\ce{Z}_0$ mapped to $\mathbb{S}_{N}$, $\ce{Z}_1$ mapped to $\mathbb{S}_N-\{\nu_1\}$, $\dots$, $\ce{Z}_{N-1}$ mapped to $\mathbb{S}_N-\{\nu_1,\nu_2,\dots,\nu_{N-1}\}$, and $\ce{Z}_N$ mapped to $\varnothing$. 

\subsection{Nonconsecutive dissociation of polyprotic acids}\label{sec:non_consecutive}

On the other hand, in non-consecutive acid dissociation, the protons can separate independently instead of consecutively. Dissociation micro-states (DMSs) are necessary to describe the non-consecutive acid dissociation. In the non-consecutive dissociation, the dissociation state $\ce{Z}_\nu$ has $\binom{N}{\nu}=\frac{N!}{(N-\nu)!\nu!}$ possible dissociation micro-states. these DMSs are given by the subsets of $\mathbb{S}_N$ with $N-\nu$ elements. The set of micro-states of $\ce{Z}_\nu$ is given by 
\begin{equation}
     M_\nu=\{ U \in\mathcal{P}( {\mathbb{S}}_N)\,:\,\left| U \right|=N-\nu\},
\end{equation}
with $\mathcal{P}( {\mathbb{S}}_N)$ as the power set of $\mathbb{S}_N$, defined as: 
\begin{equation}\label{eq:power_set}
    \mathcal{P}(\mathbb{S}_N)=\{U\,:\,U\subseteq \mathbb{S}_N \}.
\end{equation}
In simpler terms, ${M}_\nu$ represents the collection of all subsets of ${\mathbb{S}}_N$ with $N-\nu$ elements, each corresponding to a unique DMS of $\ce{Z}_\nu$. The DMSs of $\ce{Z}_\nu$ are given by $\ce{M}_\mu$ with $\mu\in {M}_\nu$. For example, the DS $\ce{Z}_1$ of a 3-protic acid has $\binom{3}{1}=3$ DMSs $\ce{M}_\mu$, with $\mu\in {M}_1=\left\{ \{1,2\},\{1,3\},\{2,3\} \right\}$. The fully protonated and fully deprotonated state $\ce{Z}_0$ and $\ce{Z}_N$ respectively, have only one DMS each $\ce{M}_{ {\mathbb{S}}_N}$ and $\ce{M}_\varnothing$, respectively.  An $N$-protic acid has a total of $2^N$ DMSs. 

The molar concentration of the deprotonation state $\ce{Z_\nu}$, $\ce{[Z_\nu]}$, is related to the molar concentrations of its deprotonation microstates, $[ \mathrm{M}_\nu]$, by
\begin{equation}\label{eq:concentrations_condition}
    \ce{[Z_\nu]}=\sum_{\mu\in  {M}_\nu}{[\mathrm{M}_\mu]},\hspace{1cm}\nu\in \mathbb{S}_N.
\end{equation}

The deprotonation and protonation micro-equilibria (DMEs and PMEs, respectively) between the DMSs $\ce{M}_\mu$ and $\ce{M}_\lambda$ are possible when the absolute difference between the sets $\mu$ and $\lambda$ is 1, and either $\mu \subsetneq \lambda$ or $\lambda \subsetneq \mu$. These micro-equilibria are given by the chemical equations
\begin{align}
    \mathrm{M}_{\mu} + \ce{ H2O <=>}& \mathrm{M}_\lambda + \ce{H3O+},\label{eq:microstates_reaction_DME}\\
    \mathrm{M}_{\lambda} + \ce{ H2O <=>}& \mathrm{M}_\mu + \ce{OH-},\label{eq:microstates_reaction_PME}
\end{align}
and the micro-equilibrium constants
\begin{align}
    K_{\mu\lambda}&=\frac{a(\mathrm{M}_\lambda) a(\ce{H3O+})}{a(\mathrm{M}_{\mu})},\label{eq:k_constant_DME}\\
    K_{\lambda\mu}&=\frac{a(\mathrm{M}_\mu) a(\ce{OH-})}{a(\mathrm{M}_{\lambda})}\label{eq:k_constant_PME}.
\end{align}
The activities of $\mathrm{M}_{\mu}$, $\mathrm{M}_{\lambda}$, $\ce{H3O+}$ and $\ce{OH-}$, are given by 
\begin{align*}
    a(\mathrm{M}_{\mu})&=\gamma_{\mu}\ce{[M_\mu]}/C^\circ,& a(\ce{M}_\lambda)&=\gamma_{\lambda}\ce{[M_\lambda]}/C^\circ,\\
    a(\ce{H3O+})&=\gamma_{\ce{H3O+}}\ce{[H3O+]}/C^\circ, \text{ and} & a(\ce{OH-})&=\gamma_{\ce{OH-}}\ce{[OH^-]}/C^\circ,
\end{align*}
respectively, with $C^\circ=1\,\mathrm{M}$ as the standard state, and $\gamma_\mu$, $\gamma_\lambda$, $\gamma_{\ce{H3O+}}$, and $\gamma_{\ce{OH-}}$, as the activity coefficients. The addition of the chemical equations \eqref{eq:microstates_reaction_DME} and \eqref{eq:microstates_reaction_PME} gives the water autoionization equilibrium
\begin{equation}
    2\ce{H2O <=> H3O+ + OH-},
\end{equation}
which has equilibrium constant 
\begin{equation}\label{eq:Kw_product}
\begin{split}
    K_\mathrm{w}&=K_{\mu\lambda}K_{\lambda\mu}\\
    &=a(\ce{H3O+})a(\ce{OH-}).
\end{split}  
\end{equation}

Although the number of micro-equilibrium constants between the DMSs of $\ce{Z}_{\nu-1}$ and $\ce{Z}_\nu$ is given by $2\binom{N}{\nu-1}\binom{N}{\nu}$, only  
$2{\binom{N}{\nu}(N-\nu)}$ are chemically related. The set of chemically related DME constants is given by 
\begin{equation}
    K=\{K_{\mu\lambda} \,:\, \mu \subsetneq \lambda \text{ or } \lambda \subsetneq \mu, \mathrm{abs}(|\mu|-|\lambda|)=1\}.
\end{equation}
The total number of DMEs and PMEs for a $N$-protic acid is given by $2^N N$. 
The 3-protic acid can be used as an example. The DMSs of $\ce{Z}_1$ and $\ce{Z}_2$ are given by $\mathrm{M}_\mu$ and $\mathrm{M}_\lambda$ respectively, with $\mu\in M_1$ and $\lambda\in M_2$ as elements of the sets  ${M}_1=\{\{2,3\},\{1,3\},\{1,2\}\}$ and $ {M}_2=\{\{1\},\{2\},\{3\}\}$, respectively. Since $\{3\}\not\subset\{1,2\}$, or $\{1,2\}\not\subset\{3\}$, there are not DMEs between the DMSs $\{1,2\}$ of $\ce{Z}_1$ and  $\{3\}$ of $\ce{Z}_2$. On the other hand, since $\{1\}\subsetneq\{1,2\}$, and $\mathrm{abs}\left(|\{1,2\}|-|\{1\}|\right)=1$, there are DMEs and PMEs between $\{1,2\}$ and $\{1\}$. In the particular case of a 3-protic acid, the total number of DMEs and PMEs is $3\times2^3=24$.

In the case of aqueous dilute solutions all the activity coefficients are approximately one, and it is justified to use the approximations $a(\mathrm{M}_\mu)\approx\left[\mathrm{M}_\mu\right]/C^\circ$, $a_{\ce{OH-}}\approx \ce{[OH^-]}/C^\circ$, and $a_{\ce{H3O+}}\approx \ce{[H3O+]}/C^\circ$. The use of the biochemical standard state $C^\standardstate=10^{-7}C^\circ$, allows to write the micro-equilibrium constants \eqref{eq:k_constant_DME}, \eqref{eq:k_constant_PME}, and \eqref{eq:Kw_product} as 
\begin{align}
    k_{\mu\lambda}&=\frac{x m_\lambda}{m_\mu},\label{eq:microequilibrium_constants_x_1}\\
    k_{\lambda\mu}&=\frac{y m_\mu}{ m_\lambda},\label{eq:microequilibrium_constants_x_2}\\
    1&=xy,\label{eq:kw_2}
\end{align}
respectively. It can be shown that $k_{\mu\lambda}=10^7 K_{\mu\lambda}$ and $k_{\lambda\mu}=10^7 K_{\lambda\mu}$.

\subsection{Relation between dissociation equilibria and micro-equilibria}\label{sec:micro-mracro-equilibrium-theo}

There is a simple relation between the deprotonation equilibrium constants $k_\nu$, with $\nu\in  \mathbb{S}_N$, and the deprotonation micro-equilibrium constants, $k_{\mu\lambda}$, with $\mu\in  {M}_{\nu-1}$, and $\lambda\in{M}_{\nu}$. The deprotonation equilibrium constant $k_\nu$,
\begin{equation}\label{eq:microequilibrium_constants_x_2.5}
    k_\nu=\frac{x z_\nu}{z_{\nu-1}},\hspace{1cm}\nu\in  \mathbb{S}_N,
\end{equation}
can be written in terms of the deprotonation micro-states
\begin{equation}\label{eq:microequilibrium_constants_x_3}
    k_\nu=\frac{x\sum_{m_\lambda \in  {M}{\nu}}m_\lambda}{\sum_{m_\mu \in  {M}{\nu-1}}m_\mu}.
\end{equation}
The reciprocal of $k_\nu$ is given by
\begin{equation}\label{eq:microequilibrium_constants_x_4}
\begin{split}
    \frac{1}{k_\nu}&=\sum_{\mu \in  {M}_{\nu-1}}\frac{m_\mu}{x\sum_{\lambda \in  {M}_{\nu}}m_\lambda}\\
    &=\sum_{\mu \in  {M}_{\nu-1}}\left({\sum_{\lambda \in  {M}_{\nu}} \frac{x m_\lambda}{m_\mu}}\right)^{-1}\\
    &=\sum_{\mu\in  {M}_{\nu-1}}\left({\sum_{\lambda \in  {M}_{\nu}}k_{\mu\lambda}}\right)^{-1}.
\end{split}
\end{equation}
A simpler relation between the equilibrium constants $k_\nu$ and the micro-equilibrium constants can be obtained by dividing numerator and denominator of the right hand side of equation \eqref{eq:microequilibrium_constants_x_3} by $m_{\mu'}$ with $\mu'\in  {M}_{\nu-1}$,
\begin{equation}\label{eq:microequilibrium_constants_x_5}
\begin{split}
    k_\nu&=\frac{x\sum_{\lambda\in  {M}_{\nu}}m_\lambda/m_{\mu'}}{\sum_{\mu \in  {M}_{\nu-1}}m_\mu/m_{\mu'}}\\
    &=\frac{\sum_{\lambda\in  {M}_{\nu}}k_{\mu'\lambda}}{\sum_{\mu\in  {M}_{\nu-1}}\tau_{\mu'\mu}},
\end{split}
\end{equation}
with $\tau_{\mu'\mu}=m_\mu/m_{\mu'}$ as the equilibrium constant for the tautomerization (isomerization) between the protonation micro-states $\mathrm{M}_\mu$ and $\mathrm{M}_{\mu'}$ of $\ce{Z}_{\nu-1}$,
\begin{equation}\label{eq:tautomerization_equilibrium}
    \mathrm{M}_{\mu'}\ce{<=>}\mathrm{M}_\mu,\hspace{1cm}\mu,{\mu'}\in  {M}_{\nu-1}.
\end{equation}
Equation \eqref{eq:microequilibrium_constants_x_5} can be written in terms of the molar fractions $x_\mu=m_\mu/z_{\nu}$ for $\mu\in M_{\nu}$,
\begin{equation}\label{eq:microequilibrium_constants_x_6}
 k_\nu=x_{\mu}\sum_{\lambda\in  {M}_{\nu}}k_{\mu\lambda},\hspace{1cm}\mu\in M_{\nu-1}.
\end{equation}
It is easy to verify that the micro-equilibrium and  tautomerization constants are related by: 
\begin{align}
    k_{\mu'\lambda}&=\tau_{\mu'\mu}k_{\mu\lambda},\label{eq:microequilibrium_constants_x_6a}\\
    k_{\mu\lambda'}&=k_{\mu\lambda}\tau_{\lambda\lambda'}.\label{eq:microequilibrium_constants_x_6b}
\end{align}

An expression for the micro-equilibrium constants, in terms of the equilibrium constants and the microscopic molar fractions, can be obtained from the general definition of the microdissociation constants, $k_{\mu\lambda}={x m_\lambda}/{m_\mu}$, for $\mu\in M_{\nu-1}$ and $\lambda\in M_\nu$. Multiplying the numerator of this expression by the unity $1={z_{\nu}}/{z_\nu}$, and the denominator by the unity $1={z_{\nu-1}}/{z_{\nu-1}}$, gives after rearranging terms
\begin{equation}
    k_{\mu\lambda}=\frac{x_\lambda}{x_\mu}k_\nu, \hspace{1cm}\mu\in M_{\nu-1}, \lambda\in M_{\nu}.
\end{equation}
This expression has been used previously, for the diprotic acid, in order to get the micro-equilibrium constants from the equilibrium constants and measurements of $\ce{^{13}C}$ NMR \cite{Laufer1984}.   

\subsection{Graph-Theory description of polyprotic acids microdissociation}\label{sec:group_theory_theo}

In graph theory, a graph $G=( {V}, {E})$ is a structure built from a set of vertices ${V}$, and a set of edges, $ {E}$. The elements of $ {E}$ are the relations between pairs of vertices. Directed and undirected edges indicate one-way and two-way relationships between two vertices, respectively. Directed edges are written in parenthesis meanwhile undirected edges are written in curly brackets. 

The dissociation of a $N$-protic acid can be represented by a graph $G_N=( {V}_N, {E}_N)$, with $ {V}_N=\{\mathrm{M}_\mu\,:\,\mu\in\mathcal{P}({\mathbb{S}}_N)\}$, and $ {E}_N$ given by the set $E_N=R_N \cup T_N$ with 
\begin{align}
    R_N&=\{\{\mathrm{M}_\mu,\mathrm{M}_\lambda\},\, \mu,\lambda\in\mathcal{P}(\mathbb{S}_N)\,:\, \mu\subsetneq\lambda \text{ or } \lambda\subsetneq\mu, \textrm{abs}(|\mu|-|\lambda|)=1\},\\
    T_N&=\{\{\mathrm{M}_\mu,\mathrm{M}_\lambda\},\, \mu,\lambda\in\mathcal{P}(\mathbb{S}_N)\,: |\mu|=|\lambda|=|\mu\cap\lambda|+1\},
\end{align}
where $R_N$ is the set of pairs of DMSs related by micro-equilibrium constants, and $T_N$ is the set of pairs of DMSs related by tautomerization constants. The set $T_N$ is the union of the tautomerization sets of each dissociation state $\mathrm{Z}_\nu$,
\begin{equation}
\begin{split}
    T_N&=\bigcup_{\nu=1}^{N-1}{T_\nu},\\
    T_\nu&=\left\{\{\mathrm{M}_\mu,\mathrm{M}_\lambda\},\mu,\lambda\in M_\nu \,:\,|\mu|=|\lambda|=|\mu\cap\lambda|+1\right\},
\end{split}
\end{equation}
with $T_\varnothing=T_{\mathbb{S}_N}=\varnothing$. 

The microequilibrium and tautomerization constants of acid dissociation always occur in pairs. For every $k_{\mu\lambda}$ and $\tau_{\mu\mu'}$, there exist $k_{\lambda \mu}$ and $\tau_{\mu'\mu}$, respectively. This pairing of constants allows undirected edges to represent pairs of equilibrium and tautomerization constants in the graph $G_N$.

A graph automorphism of $G_N$ is a permutation, denoted $\sigma$, of the set of vertices, $V_N$, that maintains the edge-vertex connectivity of $G_N$. Because the compositions of two permutations is also a permutation, the composition of two graph automorphisms must likewise be a graph automorphism. The set of automorphisms of $G_N$, equipped with the composition of automorphisms, constitutes a group known as the graph automorophism group of $G_N$, denoted $\mathrm{Aut}(G_N)$.  

\subsection{Relation between equilibrium and micro-equilibrium constants for diprotic and triprotic acids}\label{sec:results_micro_macro}

In the case of a diprotic, acid the concentration vectors of dissociation and microdissociation states are $\bm{z}^\intercal=\{z_0,z_1,z_2\}$, and
\begin{equation}
    \begin{split}
        \bm{m}^\intercal&=\{m_{\{1,2\}},m_{\{2\}}, m_{\{1\}},m_\varnothing\}\\
        &=\{m_{12},m_2,m_1,m_0\},
    \end{split}
\end{equation}
respectively. The second line of the equation defining $\bm{m}$ was obtained by applying the assignation rules $\varnothing\to 0$, $\{1\}\to 1$, $\{2\}\to 2$, and  $\{1,2\}\to 12$. In these terms, the vectors of dissociation and microdissociation constants are given by $\bm{k}^\intercal=\left\{k_1,k_2\right\}$, and $\tilde{\bm{k}}^\intercal=\{k_{12,1}, k_{12,2},k_{1,0},k_{2,0}\}$.

The use of equations \eqref{eq:microequilibrium_constants_x_5},  \eqref{eq:microequilibrium_constants_x_6a} and \eqref{eq:microequilibrium_constants_x_6b} produces the linear system of equations
\begin{align}
    k_1&=k_{12,1}+k_{12,2},\label{eq:diprotic_1}\\
    k_2(1+\tau)&=k_{12,1},\label{eq:diprotic_2}\\
    k_2(1+\tau)&=\tau k_{12,2},\label{eq:diprotic_3}\\
    0&=\tau k_{10}-k_{20}\label{eq:diprotic_4}\\
    0&=k_{12,1}-\tau k_{12,2},\label{eq:diprotic_5}
\end{align}
where $\tau=\tau_{12}$. It is easy to verify that the combined use of these equations gives the dissociation constants, $\bm{k}$, in terms of the microdissociation constants, $\tilde{\bm{k}}$. The first dissociation constant is already given by equation \eqref{eq:diprotic_1}, $k_1=k_{12,1}+k_{12,2}$. The second dissociation constant is obtained by substitution of $\tau$ from equation \eqref{eq:diprotic_5} in equation \eqref{eq:diprotic_2} to obtain
\begin{equation}\label{eq:k2_diprotic}
         k_2=\frac{k_{12,1}k_{12,2}}{k_{12,1}+k_{12,2}}.
\end{equation}
The microdissociation constants, $\tilde{\bm{k}}$, in terms of $\bm{k}$ and $\tau$ are given by
\begin{equation}\label{eq:micro_k_diprotic_2}
    \begin{split}
        k_{12,1}&=x_1 k_1,\\
        k_{12,2}&=x_2 k_1,\\
        k_{1,0}&=x_1^{-1} k_2,\\
        k_{2,0}&=x_2^{-1}k_2,
    \end{split}
\end{equation}
with $x_1=m_1/z_1$ and $x_2=m_2/z_1$.

The dissociation of the triprotic acid is a more interesting example. The concentration vector of dissociation states is $\bm{z}=\{z_0,z_1,z_2\}$. The concentration vector of the microdissociation states is 
\begin{equation}
    \bm{m}^\intercal=\{m_{ {S}_3},m_{23},m_{13},m_{12},m_3,m_2,m_1,m_0\}.
\end{equation}
The vectors of dissociation constants $\bm{k}$, and microdissociation constants $\tilde{\bm{k}}$, are given by $\bm{k}^\intercal=\{k_1,k_2,k_3\}$ and 
\begin{equation}
\begin{split}
        \tilde{\bm{k}}^\intercal=\{ & k_{S_3,23},k_{S_3,13},k_{S_3,12},k_{23,2},k_{23,3},k_{13,1},k_{13,3},k_{12,1},k_{12,2},\\
        & k_{30},k_{20},k_{10} \}.
\end{split}
\end{equation}
There are three possible tautomerizations between the dissociation micro-states $ {M}_2=\{1,2,3\}$. The use of equations \eqref{eq:microequilibrium_constants_x_6a} and \eqref{eq:microequilibrium_constants_x_6b} gives
\begin{align}
    \tau_{12}&=\frac{k_{20}}{k_{10}}=\frac{k_{12,1}}{k_{12,2}},\label{eq:tau_12_tri}\\
    \tau_{13}&=\frac{k_{30}}{k_{10}}=\frac{k_{13,1}}{k_{13,1}},\label{eq:tau_13_tri}\\
    \tau_{23}&=\frac{k_{30}}{k_{20}}=\frac{k_{23,2}}{k_{23,3}}.\label{eq:tau_23_tri}
\end{align}
In the same way, there are three tautomerizations between the dissociation micro-states $ {M}_1=\{12,13,23\}$  
\begin{align}
    \tau_{12,13}&=\frac{k_{13,1}}{k_{12,1}}=\frac{k_{S_3,12}}{k_{ {S}_3,13}},\label{eq:tau_1213_tri}\\
     \tau_{12,23}&=\frac{k_{23,2}}{k_{12,2}}=\frac{k_{S_3,12}}{k_{ {S}_3,23}},\label{eq:tau_1223_tri}\\
     \tau_{13,23}&=\frac{k_{23,3}}{k_{13,3}}=\frac{k_{S_3,13}}{k_{ {S}_3,23}}.\label{eq:tau_1323_tri}
\end{align}
The tautomerization constants are related by
 \begin{align}
     \tau_{13}&=\tau_{12}\tau_{23},\\
     \tau_{12,23}&=\tau_{12,13}\tau_{13,23}.
 \end{align}
 
The use of equation \eqref{eq:microequilibrium_constants_x_5} with $\nu=1$ gives the first dissociation constant of the triprotic acid $k_1=k_{S_3,1}+k_{S_3,2}+k_{S_3,3}$. Equation \eqref{eq:microequilibrium_constants_x_5} with $\nu=2$ gives three equations for $k_2$, which are all equivalent to
\begin{equation}\label{eq:k2_triprotic_0}
    (1+\tau_{12}+\tau_{13})k_2=k_{12,1}+k_{13,1}+\tau_{12}k_{23,2}.
\end{equation}
The use of the tautomerizations \eqref{eq:tau_12_tri}-\eqref{eq:tau_23_tri}, in terms of the dissociation micro-states of $ \ce{Z}_1$, gives
\begin{equation}\label{eq:k2_triprotic}
    k_2=\frac{k_{13,3}\left(k_{12,1}k_{12,2}+k_{13,1}k_{12,2}+k_{12,1}k_{23,2}\right)}{k_{13,1}k_{12,2}+k_{12,1}k_{13,3}+k_{12,2}k_{13,3}}.
\end{equation}
This is only one of three possible (equivalent) equations for $k_2$. The third dissociation constant of the triprotic acid gives also three equations equivalent to
\begin{equation}\label{eq:k3_triprotic_0}
    (1+\tau_{12,13}+\tau_{12,23})k_3=k_{{S}_3,12}.
\end{equation}
This equation gives $k_{{S}_3,12}=x_{12}^{-1}k_3$. The use of the tautomerizations \eqref{eq:tau_1213_tri}-\eqref{eq:tau_1323_tri}, in terms of the dissociation micro-states of $S_2$, in equation \eqref{eq:k3_triprotic_0} gives
\begin{equation}\label{eq:k3_triprotic}
    k_3=\frac{k_{{S}_3,12}k_{{S}_3,13}k_{{S}_3,23}}{k_{{S}_3,12}k_{{S}_3,13}+k_{{S}_3,12}k_{{S}_3,23}+k_{{S}_3,13}k_{{S}_3,23}}.
\end{equation}
Again, this is only one of three possible (equivalent) equations for $k_3$.

The use of the tautomerizations \eqref{eq:tau_13_tri} and\eqref{eq:tau_23_tri} in $k_1=k_{S_3,1}+k_{s_3,2}+k_{S_3,3}$ gives 
\begin{equation}\label{eq:k01_triprotic}
    k_{S_3,1}=\frac{k_1}{1+\tau_{12}+\tau_{13}}=x_1 k_1. 
\end{equation}
The constants $k_{S_3,2}$ and $k_{S_3,3}$ are obtained from equations \eqref{eq:tau_12_tri} and \eqref{eq:tau_13_tri}
\begin{align}
    k_{S_3,2}&=\tau_{12}k_{S_3,1}=x_2 k_1,\\
    k_{S_3,3}&=\tau_{13}k_{S_3,1}=x_3 k_1.    
\end{align}

The constant $k_{12,1}$ is obtained from equation \eqref{eq:k2_triprotic_0} and the tautomerizations. The substitutions $k_{13,1}=\tau_{12,13}k_{12,1}$ and $\tau_{12}k_{23,2}=\tau_{12,23}k_{12,1}$ in \eqref{eq:k2_triprotic_0} gives
\begin{equation}\label{eq:k112_triprotic}
    k_{12,1}=\frac{1+\tau_{12}+\tau_{13}}{1+\tau_{12,13}+\tau_{12,23}}k_2=\frac{x_{12}}{x_1}k_2.
\end{equation}
The other constants of $\tilde{\bm{k}}$ are obtained from $k_{12,1}$ and the tautomerizations
\begin{align}
    k_{13,1}&=\tau_{12,13}k_{12,1}=\frac{x_{13}}{x_1}k_2,\\
    k_{12,2}&=\frac{k_{12,1}}{\tau_{12}}=\frac{x_{12}}{x_2}k_2,\\
    k_{23,2}&=\frac{\tau_{12,23}}{\tau_{12}}k_{12,1}=\frac{x_{23}}{x_2}k_2,\\
    k_{13,3}&=\frac{\tau_{12,13}}{\tau_{13}}k_{12,1}=\frac{x_{13}}{x_3}k_2,\\
    k_{23,3}&=\frac{\tau_{12,23}}{\tau_{13}}k_{12,1}=\frac{x_{23}}{x_3}k_2.
\end{align}

Finally, the constants $k_{{S}_3,\mu}$, $\mu\in M_2$, are given by equation \eqref{eq:k3_triprotic_0} and
\begin{align}
    k_{{S}_3,13}&=\tfrac{1}{\tau_{12,13}}k_{{S}_3,12}=x_{13}^{-1}k_3,\\
    k_{{S}_3,23}&=\tfrac{1}{\tau_{12,23}}k_{{S}_3,12}=x_{23}^{-1}k_3.
\end{align}

\subsection{Graph automorphism groups of the dissociation of polyprotic acids}\label{sec:results_graph_automorphism}

\begin{figure}
    \centering
\begin{tikzpicture}

    \Vertex[label=$ \mathrm{M}_1$, x=0, y=4.25, opacity=0.2,color=black]{A11} 
    \Vertex[label=$ \mathrm{M}_0$,x=2,y=4.25, opacity=0.2, color=black]{B11}
    \Edge[style=red](A11)(B11)
    \Text[x=2.5,y=5.25]{(a)}

    \Vertex[label=${ \mathrm{M}_{12}}$,x=4, y=4.25,opacity=0.2, color=black]{A22} \Vertex[label=${ \mathrm{M}_2}$,x=6,y=5.25,opacity=0.2,color=black]{B22}
    \Vertex[label=${ \mathrm{M}_1}$,x=6,y=3.25,opacity=0.2, color=black]{C22}
    \Vertex[label=${ \mathrm{M}_0}$,x=8,y=4.25,opacity=0.2, color=black]{D22}
    \Edge[style=red,fontcolor=black](A22)(B22) 
    \Edge[style=green,fontcolor=black](A22)(C22) 
    \Edge[style=gray,fontcolor=black](B22)(C22) 
    \Edge[style=green,fontcolor=black](B22)(D22) 
    \Edge[style=red,fontcolor=black](C22)(D22)
    \Text[x=8.5,y=5.25]{(b)}


    \Vertex[label=${ \mathrm{M}_{ {S}_3}}$,opacity=0.2,color=black]{A} \Vertex[label=${ \mathrm{M}_{23}}$,x=2,y=2,opacity=0.2,color=black]{B1}
    \Vertex[label=${ \mathrm{M}_{13}}$,x=3,y=0,opacity=0.2,color=black]{B2}
    \Vertex[label=${ \mathrm{M}_{12}}$,x=2,y=-2,opacity=0.2,color=black]{B3}
    \Vertex[label=${ \mathrm{M}_3}$,x=6,y=2,opacity=0.2,color=black]{C1}
    \Vertex[label=$ \mathrm{M}_2$,x=5,y=0,opacity=0.2,color=black]{C2}
    \Vertex[label=$ \mathrm{M}_1$,x=6,y=-2,opacity=0.2,color=black]{C3}
    \Vertex[label=$ \mathrm{M}_0$,x=8,y=0,opacity=0.2,color=black]{D}

    \Edge[style=red](A)(B1)
    \Edge[style={gray,dashed}](B1)(B3)
    \Edge[style=green](A)(B2)
    \Edge[style=blue](A)(B3)
    \Edge[style=gray](B1)(B2)
    \Edge[style=gray](B2)(B3)

    \Edge[style=green](B1)(C1)
    \Edge[style={blue,dashed}](B1)(C2)
    \Edge[style=red](B2)(C1)

    \Edge[style={red,dashed}](B3)(C2)
    \Edge[style=blue](B2)(C3)
    \Edge[style=green](B3)(C3) 

    \Edge[style={gray,dashed}](C1)(C2)
    \Edge[style={green,dashed}](C2)(D)
    \Edge[style=gray](C1)(C3)
    \Edge[style={gray,dashed}](C2)(C3)
    
    \Edge[style=blue](C1)(D)
    \Edge[style=red](C3)(D)
    \Text[x=8.5,y=1]{(c)}

    \end{tikzpicture}
    \caption{Dissociation graphs of some polyprotic acids. (a) Monoprotic acid; (b) Diprotic acid; (c) Triprotic acid. Red, green and blue edges represent the deprotonation/protonation of protons (1), (2), and (3), respectively, gray edges represent tautomerizations. Dashed edges are used to facilitate the visualization of $G_3$.}
    \label{fig:polyprotic_graphs}
\end{figure}

In the case of a monoprotic acid, the sets $ {V}_1$ and $ {E}_1$ are $ {V}_1=\{\mathrm{M}_1,\mathrm{M}_0\}$ and $ {E}_1=\{\{\mathrm{M}_1,\mathrm{M}_0\}\}$. The monoprotic acid does not display tautomerizations. The graph $G_1=( {V}_1, {E}_1)$ is depicted in Figure \ref{fig:polyprotic_graphs}(a). This graph is known as the complete graph of order two, denoted as $K_2$, where all vertices are connected by edges. The acid-base permutation
\begin{equation}\label{eq:acid-base_permutation_monoprotic}
    \sigma_{10}=(\mathrm{M}_1 \mathrm{M}_0)(\ce{H3O+}\ce{OH-}),
\end{equation}
exchanges $\mathrm{M}_1$ with $\mathrm{M}_0$, and $\ce{H3O+}$ with $\ce{OH-}$. In terms of concentrations, $\sigma_{10}$ exchanges $m_1$ with $m_0$, and $x$ with $y$. The effect of $\sigma_{10}$ on the micro-equilibrium constants is given by $\sigma_{10}(k_{10})=k_{01}$ and $\sigma_{10}(k_{01})=k_{10}$. The graph $G_1$ is shown to be preserved under the action of $\sigma_{10}$ in Figure \ref{fig:sigma_monoprotic}. This means  $\sigma_{10}$ maps $G_1$ onto itself without losing edge connectivity, making $\sigma_{10}$ a graph automorphism of $G_1$. The identity permutation $e=(\mathrm{M}_1)(\mathrm{M}_0)(\ce{H3O+})(\ce{OH-})$ is also a graph automorphism of $G_1$. Under the operation of composition, the graph automorphisms $e$ and $\sigma_{10}$ generate the graph automorphism group of $G_1$, denoted as $\mathrm{Aut}(G_1)=\left<\sigma_{10} \right>$. The permutation $\sigma_{10}$ is an involution, meaning $\sigma_{10}=\sigma_{10}^{-1}$, hence $\sigma_{10}^2=e$ establishes a condition on the generators of the group. The group presentation is given as 
\begin{equation}
\mathrm{Aut}(G_1)=\left< \sigma_{10} \mid \sigma_{10}^2=e \right>,    
\end{equation}
which specifies that the cyclic group of order two, $C_2$, represents the monoprotic acid dissociation.

\begin{figure}
    \centering
    \begin{tikzpicture}
        \Vertex[label=$ \mathrm{M}_{1}$,x=0,y=0,opacity=0.2,color=black]{M1}
        \Vertex[label=$ \mathrm{M}_{0}$,x=0,y=2,opacity=0.2,color=black]{M2}
        \Vertex[Pseudo,x=0.,y=0.75,color=white]{Mr1}
        \Vertex[Pseudo,x=0.,y=1.25,color=white]{Mr2}
        \Edge[style=red](M1)(M2)
        \Vertex[label=$ \mathrm{M}_{0}$,x=2,y=0,opacity=0.2,color=black]{M0a}
        \Vertex[label=$ \mathrm{M}_{1}$,x=2,y=2,opacity=0.2,color=black]{M1a}
        \Vertex[Pseudo,x=2,y=0.75,color=white]{Mla1}
        \Vertex[Pseudo,x=2,y=1.25,color=white]{Mla2}
        \Edge[style=red](M0a)(M1a)

    \Edge[Direct,lw=2.5,color=blue!60](Mr1)(Mla1)
    \Text[x=1,y=0.45]{$\sigma_{10}$}
    \Edge[Direct,lw=2.5,color=blue!60](Mla2)(Mr2)
    \Text[x=1,y=1.55]{$\sigma_{10}$}
    
    \end{tikzpicture}
    \caption{Effect of the permutations $\sigma_{10}$ on the graph $G_1$.}
    \label{fig:sigma_monoprotic}
\end{figure}

The dissociation of a diprotic acid is represented by the graph $G_2=(V_2,E_2)$ with
\begin{align}
    {V}_2&=\{\mathrm{M}_{12},\mathrm{M}_2,\mathrm{M}_1,\mathrm{M}_0\},\\
    {E}_2&=\{\{\mathrm{M}_{12},\mathrm{M}_{2}\}, \{\mathrm{M}_{12},\mathrm{M}_1\},\{\mathrm{M}_1,\mathrm{M}_2\}, \{\mathrm{M}_2,\mathrm{M}_0\},\{\mathrm{M}_1,\mathrm{M}_0\}\}.
\end{align}
The edge set is the union $E_2=R_2 \cup T_2$ of the microdissociation set $R_2$, and the tautomerization set $T_2=\left\{ \left\{ \mathrm{M}_1,\mathrm{M}_2 \right\} \right\}$. The graph $G_2=( {V}_2, {E}_2)$ of the diprotic acid microdissociations is depicted in Figure \ref{fig:polyprotic_graphs}(b). In this graph, red and green edges represent two types of microdissociations. The red edges correspond to the dissociations of protons occupying site 1, while the green edges correspond to protons at site 2. The gray edge represents the tautomerization between the DMSs $\mathrm{M}_1$ and $\mathrm{M}_2$. The graph $G_2$ of the diprotic acid microdissociations is a complete tripartite graph, $G_2=K_{1,1,2}$, also known as the diamond graph. The acid-base permutation $\sigma_{12,0}$, and the tautomerization permutation $\sigma_{21}$ are defined as:
\begin{align}
    \sigma_{12,0}&=(\mathrm{M}_{12} \mathrm{M}_0)(\ce{H3O+}\ce{OH-}),\label{eq:acid-base_permutation_diprotic}\\
    \sigma_{21}&=(\mathrm{M}_2 \mathrm{M}_1)(\ce{H3O+})(\ce{OH-}).\label{eq:tautomerization_permutation_diprotic}
\end{align}
These permutations preserve the edge-vertex connectivity of $G_2$, as shown in Figure \ref{fig:sigma_rho_diprotic}. Under the composition operation,  the graph automorphisms $\sigma_{12,0}$, $\sigma_{21}$, and the identity $e$, form the graph automorphism group of $G_2$, denoted $\mathrm{Aut}(G_2)=\left<\sigma_{12,0},\sigma_{21} \right>$. Since the permutations $\sigma_{12,0}$ and $\sigma_{21}$ are involutions (self-inverses), the graph automorphism group of $G_2$ is given by 
\begin{equation}
    \mathrm{Aut}(G_2)=\left<\sigma_{12,0},\sigma_{21}\mid \sigma_{12,0}^2=\sigma_{21}^2=(\sigma_{12,0}\sigma_{21})^2=e\right>.
\end{equation}
This is known as Klein's 4-group, or $C_2 \times C_2$. In Figure \ref{fig:sigma_rho_diprotic}, it is evident that the product $\rho=\sigma_{12,0}\sigma_{21}$ results in a counterclockwise rotation of $G_2$ by $\pi$ radians.  Since $C_2$ and $S_2$ are isomorphic ($C_2 \cong S_2$), the automorphism group of the graph representing the diprotic acid is also expressed as $\mathrm{Aut}(G_2)=C_2\times S_2$.

\begin{figure}
    \centering
    \begin{tikzpicture}
        \Vertex[label=$ \mathrm{M}_{12}$,x=-2,y=4,opacity=0.2,color=black]{M12}
        \Vertex[label=$ \mathrm{M}_{2}$,x=-2,y=6,opacity=0.2,color=black]{M2}
        \Vertex[label=$ \mathrm{M}_{1}$,x=0,y=4,opacity=0.2,color=black]{M1}
        \Vertex[label=$ \mathrm{M}_{0}$,x=0,y=6,opacity=0.2,color=black]{M0}
        \Vertex[Pseudo,x=-0.85,y=2.,color=white]{Mu1}
        \Vertex[Pseudo,x=-1.15,y=2.,color=white]{Mu2}
        \Vertex[Pseudo,x=0.,y=0.85,color=white]{Mr1}
        \Vertex[Pseudo,x=0.,y=1.15,color=white]{Mr2}
        \Edge[style=red](M12)(M2)
        \Edge[style=green](M12)(M1)
        \Edge[style=green](M2)(M0)
        \Edge[style=red](M1)(M0)
        \Edge[style=gray](M2)(M1)

        \Vertex[label=$ \mathrm{M}_{0}$,x=-2 ,y=0, opacity=0.2,color=black]{M12a}
        \Vertex[label=$ \mathrm{M}_{2}$,x=-2,y=2,opacity=0.2,color=black]{M2a}
        \Vertex[label=$ \mathrm{M}_{1}$,x=0,y=0,opacity=0.2,color=black]{M1a}
        \Vertex[label=$ \mathrm{M}_{12}$,x=0,y=2,opacity=0.2,color=black]{M0a}
        \Vertex[Pseudo,x=-1.15,y=4,color=white]{Mda1}
        \Vertex[Pseudo,x=-0.85,y=4,color=white]{Mda2}
        \Vertex[Pseudo,x=0,y=4.85,color=white]{Mra1}
        \Vertex[Pseudo,x=0,y=5.15,color=white]{Mra2}
        \Edge[style=green](M12a)(M2a)
        \Edge[style=red](M12a)(M1a)
        \Edge[style=red](M2a)(M0a)
        \Edge[style=green](M1a)(M0a)
        \Edge[style=gray](M2a)(M1a)
        
        \Vertex[label=$ \mathrm{M}_{12}$,x=2 ,y=4, opacity=0.2,color=black]{M12b}
        \Vertex[label=$ \mathrm{M}_{1}$,x=2,y=6,opacity=0.2,color=black]{M2b}
        \Vertex[label=$ \mathrm{M}_{2}$,x=4,y=4,opacity=0.2,color=black]{M1b}
        \Vertex[label=$ \mathrm{M}_{0}$,x=4,y=6,opacity=0.2,color=black]{M0b}
        \Vertex[Pseudo,x=2.85,y=2.,color=white]{Mub1}
        \Vertex[Pseudo,x=3.15,y=2.,color=white]{Mub2}
        \Vertex[Pseudo,x=2,y=0.85,color=white]{Mlb1}
        \Vertex[Pseudo,x=2,y=1.15,color=white]{Mlb2}
        \Edge[style=green](M12b)(M2b)
        \Edge[style=red](M12b)(M1b)
        \Edge[style=red](M2b)(M0b)
        \Edge[style=green](M1b)(M0b)
        \Edge[style=gray](M2b)(M1b) 

        \Vertex[label=$ \mathrm{M}_{0}$,x=2 ,y=0, opacity=0.2,color=black]{M12b}
        \Vertex[label=$ \mathrm{M}_{1}$,x=2,y=2,opacity=0.2,color=black]{M2b}
        \Vertex[label=$ \mathrm{M}_{2}$,x=4,y=0,opacity=0.2,color=black]{M1b}
        \Vertex[label=$ \mathrm{M}_{12}$,x=4,y=2,opacity=0.2,color=black]{M0b}
        \Vertex[Pseudo,x=2.85,y=4,color=white]{Mdc1}
        \Vertex[Pseudo,x=3.15,y=4,color=white]{Mdc2}
        \Vertex[Pseudo,x=2,y=4.85,color=white]{Mlc1}
        \Vertex[Pseudo,x=2,y=5.15,color=white]{Mlc2}
        \Edge[style=red](M12b)(M2b)
        \Edge[style=green](M12b)(M1b)
        \Edge[style=green](M2b)(M0b)
        \Edge[style=red](M1b)(M0b)
        \Edge[style=gray](M2b)(M1b) 

    \Edge[Direct,lw=2.5,color=blue!60](Mda1)(Mu2)
    \Edge[Direct,lw=2.5,color=blue!60](Mu1)(Mda2)
    \Text[x=-1.4,y=3,rotation=90]{$\sigma_{12,0}$}
    \Text[x=-0.6,y=3,rotation=-90]{$\sigma_{12,0}$}
    \Edge[Direct,lw=2.5,color=gray!60](Mr1)(Mlb1)
    \Edge[Direct,lw=2.5,color=gray!60](Mlb2)(Mr2)
    \Text[x=1,y=0.6]{$\sigma_{21}$}
    \Text[x=1,y=1.4]{$\sigma_{21}$}
    \Edge[Direct,lw=2.5,color=blue!69](Mdc1)(Mub1)
    \Edge[Direct,lw=2.5,color=blue!69](Mub2)(Mdc2)
    \Text[x=2.6,y=3,rotation=90]{$\sigma_{12,0}$}
    \Text[x=3.4,y=3,rotation=-90]{$\sigma_{12,0}$}
    \Edge[Direct,lw=2.5,color=gray!60](Mra1)(Mlc1)
    \Edge[Direct,lw=2.5,color=gray!60](Mlc2)(Mra2)
    \Text[x=1,y=5.4]{$\sigma_{21}$} 
    \Text[x=1,y=4.6]{$\sigma_{21}$}
    
    \end{tikzpicture}
    \caption{Effect of the permutations $\sigma_{12,0}$ and $\sigma_{21}$ on the graph $G_2$.}
    \label{fig:sigma_rho_diprotic}
\end{figure}

The triprotic acid dissociation is characterized by the sets $V_3$ and $E_3$ 
\begin{align}
    {V}_3&=\{\mathrm{M}_\mu\,:\,\mu\in\mathcal{P}({\mathbb{S}}_3)\},\\
    E_3&= R_3 \cup T_3,
\end{align}
where $R_3$ is the set of pairs of DMSs related by micro-equilibrium constants, and $T_3$ is the set of pairs of DMSs related by tautomerization constants. The graph $G_3=( {V}_3, {E}_3)$ is shown in Figure \ref{fig:polyprotic_graphs}(c). This graph displays three types of edges representing deprotonations: proton 1 in red, proton 2 in green, and proton 3 in blue. Tautomerizations are represented by gray edges. There are three permutations that serve as generators of the graph automorphism group of $G_3$:
\begin{align}
    \sigma_{\mathrm{d}}&=\sigma_{12,23}\sigma_{1,2}\nonumber\\
    &=(\mathrm{M}_{12},\mathrm{M}_{23})(\mathrm{M}_1,\mathrm{M}_2),\label{eq:sigma_triprotic_1}\\
    \sigma_{\mathrm{d}}'&=\sigma_{12,13}\sigma_{2,3}\nonumber\\
    &=(\mathrm{M}_{12},\mathrm{M}_{13})(\mathrm{M}_2,\mathrm{M}_3),\label{eq:sigma_triprotic_2}\\
    c_2&=\sigma_{\mathbb{S}_3,0}\sigma_{12,1}\sigma_{13,2}\sigma_{23,3}\nonumber\\
    &=(\mathrm{M}_{\mathbb{S}_3},\mathrm{M}_0)(\mathrm{M}_{12},\mathrm{M}_{13})(\mathrm{M}_{13},\mathrm{M}_2)(\mathrm{M}_{23},\mathrm{M}_3)(\ce{H3O+}\ce{OH-}),\label{eq:sigma_triprotic_3}
\end{align}
with $\sigma_{\mathrm{d}}$ and $\sigma_{\mathrm{d}}'$ as tautomerization permutations, and $c_2$ as an acid-base permutation. The presentation of this group is given by
\begin{equation}\label{eq:graph_aut_group_G3}
    \mathrm{Aut}(G_3)=\left<\sigma_{\mathrm{d}},\sigma_{\mathrm{d}}',c_2\mid\sigma_{\mathrm{d}}^2=(\sigma_{\mathrm{d}}')^2=c_2^2=e\right>.
\end{equation}
The effect of these permutations on $G_3$ is depicted in Figure \ref{fig:effect_sigmas_G3}. The permutations \eqref{eq:sigma_triprotic_1}-\eqref{eq:sigma_triprotic_3} can be interpreted as symmetry operations:

\begin{itemize}
    \item The tautomerization $\sigma_{\mathrm{d}}$ acts as a reflection on the plane containing $\mathrm{M}_{\mathbb{S}_3}$, $\mathrm{M}_{12}$, $\mathrm{M}_{3}$, and $\mathrm{M}_{0}$.
    \item The tautomerization $\sigma_{\mathrm{d}}'$ is a reflection on the plane containing $\mathrm{M}_{S_3}$, $\mathrm{M}_{23}$, $\mathrm{M}_{1}$, and $\mathrm{M}_{0}$.
    \item The acid-base permutation $c_2$ is a rotation about the axis that goes from the center of the graph through the midpoint between the DMSs $\mathrm{M}_{23}$ and $\mathrm{M}_{3}$.
\end{itemize}

\begin{figure}
    \centering
    \begin{tikzpicture}
    \Vertex[label=${ \mathrm{M}_{{S}_3}}$,x=-3,y=1, opacity=0.2,color=black]{MS3a} 
    \Vertex[label=${ \mathrm{M}_{23}}$,x=-2,y=2,opacity=0.2,color=black]{M23a}
    \Vertex[label=${ \mathrm{M}_{13}}$,x=-1.5,y=1,opacity=0.2,color=black]{M13a}
    \Vertex[label=${ \mathrm{M}_{12}}$,x=-2,y=0,opacity=0.2,color=black]{M12a}
    \Vertex[label=${ \mathrm{M}_3}$,x=0.,y=2,opacity=0.2,color=black]{M3a}
    \Vertex[label=$ \mathrm{M}_2$,x=-0.5,y=1,opacity=0.2,color=black]{M2a}
    \Vertex[label=$ \mathrm{M}_1$,x=0.,y=0,opacity=0.2,color=black]{M1a}
    \Vertex[label=$ \mathrm{M}_0$,x=1,y=1,opacity=0.2,color=black]{M0a}
    
    \Vertex[Pseudo,x=-1,y=0]{downa1}
    \Vertex[Pseudo,x=0.5,y=0.5]{downa2}
    \Vertex[Pseudo,x=0.5,y=1.75]{downa3}

    \Edge[style=red](MS3a)(M23a)
    \Edge[style={gray,dashed}](M23a)(M12a)
    \Edge[style=green](MS3a)(M13a)
    \Edge[style=blue](MS3a)(M12a)
    \Edge[style=gray](M23a)(M13a)
    \Edge[style=gray](M13a)(M12a)

    \Edge[style=green](M23a)(M3a)
    \Edge[style={blue,dashed}](M23a)(M2a)
    \Edge[style=red](M13a)(M3a)

    \Edge[style={red,dashed}](M12a)(M2a)
    \Edge[style=blue](M13a)(M1a)
    \Edge[style=green](M12a)(M1a) 

    \Edge[style={gray,dashed}](M2a)(M3a)
    \Edge[style={green,dashed}](M2a)(M0a)
    \Edge[style=gray](M3a)(M1a)
    \Edge[style={gray,dashed}](M2a)(M1a)
   
    \Edge[style=blue](M3a)(M0a)
    \Edge[style=red](M1a)(M0a)
    \Vertex[label=${ \mathrm{M}_{{S}_3}}$, x=-3,y=-3,opacity=0.2,color=black]{MS3b} 
    \Vertex[label=${ \mathrm{M}_{23}}$,x=-2,y=-2,opacity=0.2,color=black]{M23b}
    \Vertex[label=${ \mathrm{M}_{12}}$,x=-1.5,y=-3,opacity=0.2,color=black]{M12b}
    \Vertex[label=${ \mathrm{M}_{13}}$,x=-2,y=-4,opacity=0.2,color=black]{M13b}
    \Vertex[label=${ \mathrm{M}_2}$,x=0,y=-2,opacity=0.2,color=black]{M2b}
    \Vertex[label=$ \mathrm{M}_3$,x=-0.5,y=-3,opacity=0.2,color=black]{M3b}
    \Vertex[label=$ \mathrm{M}_1$,x=0,y=-4,opacity=0.2,color=black]{M1b}
    \Vertex[label=$ \mathrm{M}_0$,x=1,y=-3,opacity=0.2,color=black]{M0b}

    \Vertex[Pseudo,x=-1,y=-2]{upb}

    \Edge[style=red](MS3b)(M23b)
    \Edge[style={gray}](M23b)(M12b)
    \Edge[style=green](MS3b)(M13b)
    
    \Edge[style={gray,dashed}](M23b)(M13b)
    \Edge[style=gray](M13b)(M12b)
    \Edge[style=blue](MS3b)(M12b)

    \Edge[style={green,dashed}](M23b)(M3b)
    \Edge[style={blue}](M23b)(M2b)
    \Edge[style={red,dashed}](M13b)(M3b)
    \Edge[style={red}](M12b)(M2b)
    \Edge[style=blue](M13b)(M1b)
    \Edge[style=green](M12b)(M1b) 
    \Edge[style={gray,dashed}](M2b)(M3b)
    \Edge[style={green}](M2b)(M0b)
    \Edge[style={gray,dashed}](M3b)(M1b)
    \Edge[style={blue,dashed}](M3b)(M0b)
    \Edge[style={gray}](M2b)(M1b)
    
    \Edge[style=red](M1b)(M0b)


    \Vertex[label=${ \mathrm{M}_{{S}_3}}$, x=3,y=1,opacity=0.2,color=black]{MS3c} 
    \Vertex[label=${ \mathrm{M}_{13}}$,x=4,y=2,opacity=0.2,color=black]{M13c}
    \Vertex[label=${ \mathrm{M}_{23}}$,x=4.5,y=1,opacity=0.2,color=black]{M23c}
    \Vertex[label=${ \mathrm{M}_{12}}$,x=4,y=0,opacity=0.2,color=black]{M12c}
    \Vertex[label=${ \mathrm{M}_3}$,x=6,y=2,opacity=0.2,color=black]{M3c}
    \Vertex[label=$ \mathrm{M}_1$,x=5.5,y=1,opacity=0.2,color=black]{M1c}

    \Vertex[label=$ \mathrm{M}_2$,x=6,y=0,opacity=0.2,color=black]{M2c}
    \Vertex[label=$ \mathrm{M}_0$,x=7,y=1,opacity=0.2,color=black]{M0c}
    \Vertex[Pseudo,x=3.5,y=-2.5]{leftc}

    \Edge[style={blue}](MS3c)(M12c)
    
    \Edge[style={gray,dashed}](M13c)(M12c)
    \Edge[style=green](MS3c)(M13c)
    \Edge[style={gray}](M23c)(M13c)
    \Edge[style=gray](M23c)(M12c)
    \Edge[style=red](MS3c)(M23c) 
    
    \Edge[style={red}](M13c)(M3c)
    \Edge[style={red}](M12c)(M2c)
    \Edge[style={blue,dashed}](M13c)(M1c)
    \Edge[style={green,dashed}](M12c)(M1c) 
    \Edge[style={blue}](M23c)(M2c)
    \Edge[style={green}](M23c)(M3c)

    \Edge[style={green}](M2c)(M0c)
    \Edge[style={gray,dashed}](M3c)(M1c)
    
    \Edge[style={blue}](M3c)(M0c)
    \Edge[style={gray,dashed}](M2c)(M1c)
    \Edge[style={red,dashed}](M1c)(M0c)
    \Edge[style={gray}](M2c)(M3c)
    \Vertex[label=$ \mathrm{M}_{0}$,x=3,y=-3,opacity=0.2,color=black]{M0d} 
    \Vertex[label=${ \mathrm{M}_{3}}$,x=4,y=-2,opacity=0.2,color=black]{M3d}
    \Vertex[label=${ \mathrm{M}_{2}}$,x=4.5,y=-3,opacity=0.2,color=black]{M2d}
    \Vertex[label=${ \mathrm{M}_{1}}$,x=4,y=-4,opacity=0.2,color=black]{M1d}
    \Vertex[label=${ \mathrm{M}_{23}}$,x=6,y=-2,opacity=0.2,color=black]{M23d}
    \Vertex[label=$ \mathrm{M}_{13}$,x=5.5,y=-3,opacity=0.2,color=black]{M13d}
    \Vertex[label=$ \mathrm{M}_{12}$,x=6,y=-4,opacity=0.2,color=black]{M12d}
    \Vertex[label=$ \mathrm{M}_{S_3}$,x=7,y=-3,opacity=0.2,color=black]{MS3d}

    \Vertex[Pseudo,x=3.5,y=1.75]{leftd}

    \Edge[style=red](MS3d)(M23d)
        \Edge[style={gray,dashed}](M23d)(M13d)
        \Edge[style={green,dashed}](MS3d)(M13d)
            \Edge[style={gray}](M23d)(M12d)
    \Edge[style={blue}](MS3d)(M12d)
    \Edge[style={gray,dashed}](M13d)(M12d)

    \Edge[style={blue,dashed}](M13d)(M1d)
    \Edge[style={green}](M12d)(M1d)
  
    \Edge[style={red}](M12d)(M2d)
    \Edge[style={red,dashed}](M13d)(M3d)
          \Edge[style={blue}](M23d)(M2d)
    \Edge[style={green}](M23d)(M3d)
    \Edge[style={gray}](M2d)(M3d)
    \Edge[style={gray,dashed}](M3d)(M1d)
    \Edge[style={green}](M2d)(M0d)
    \Edge[style={gray}](M2d)(M1d)    
    \Edge[style={blue}](M3d)(M0d)

    \Edge[style=red](M1d)(M0d)
    \Edge[Direct, style=green!60,lw=2.5](downa1)(upb)
    \Text[x=-1.9,y=-1]{$\sigma_{12,13}\sigma_{2,3}$}
    \Edge[Direct, style=red!60,lw=2.5](downa3)(leftd)
    \Text[x=2,y=2.1]{$\sigma_{13,23}\sigma_{1,2}$}
    \Edge[Direct, style=blue!60,lw=2.5](downa2)(leftc)
    \Text[x=2.2,y=-0.7,rotation=-45]{$\sigma_{S_3,0}\sigma_{12,1}\sigma_{13,2}\sigma_{23,3}$}
    \end{tikzpicture}
    \caption{Effect of the permutations $\sigma_{13,23}\sigma_{1,2}$, $\sigma_{12,13}\sigma_{2,3}$, and $\sigma_{S_3,0}\sigma_{12,1}\sigma_{13,2}\sigma_{23,3}$, on the graph $G_3$.}
    \label{fig:effect_sigmas_G3}
\end{figure}

The other graph automorphisms of $G_3$ are
\begin{align}
    \sigma_\mathrm{d}''&=(\mathrm{M}_{12},\mathrm{M}_{23})(\mathrm{M}_{1},\mathrm{M}_{3}),\label{eq:sigma_triprotic_4}\\
    c_3&=(\mathrm{M}_{12},\mathrm{M}_{13},\mathrm{M}_{23})(\mathrm{M}_{1},\mathrm{M}_{3},\mathrm{M}_{2}),\label{eq:sigma_triprotic_5}\\   
    c_3^2&=(\mathrm{M}_{12},\mathrm{M}_{23},\mathrm{M}_{13})(\mathrm{M}_{1},\mathrm{M}_{2},\mathrm{M}_{3}),\label{eq:sigma_triprotic_6}\\ 
    c_2'&=(\mathrm{M}_{0},\mathrm{M}_{S_3})(\mathrm{M}_{12},\mathrm{M}_{2})(\mathrm{M}_{13},\mathrm{M}_{3})(\mathrm{M}_{23},\mathrm{M}_{1})(\ce{H3O+}\ce{OH-}),\label{eq:sigma_triprotic_7}\\
    c_2''&=(\mathrm{M}_{0},\mathrm{M}_{S_3})(\mathrm{M}_{12},\mathrm{M}_{3})(\mathrm{M}_{13},\mathrm{M}_{1})(\mathrm{M}_{23},\mathrm{M}_{2})(\ce{H3O+}\ce{OH-}),\label{eq:sigma_triprotic_8}\\
    \mathrm{i}&=(\mathrm{M}_{0},\mathrm{M}_{S_3})(\mathrm{M}_{12},\mathrm{M}_{3})(\mathrm{M}_{13},\mathrm{M}_{2})(\mathrm{M}_{23},\mathrm{M}_{1})(\ce{H3O+}\ce{OH-}),\label{eq:sigma_triprotic_9}\\
    s_6&=(\mathrm{M}_{0},\mathrm{M}_{S_3})(\mathrm{M}_{12},\mathrm{M}_{1},\mathrm{M}_{13},\mathrm{M}_{3},\mathrm{M}_{23},\mathrm{M}_{2})(\ce{H3O+}\ce{OH-}),\label{eq:sigma_triprotic_10}\\
    s_6'&=(\mathrm{M}_{0},\mathrm{M}_{S_3})(\mathrm{M}_{12},\mathrm{M}_{2},\mathrm{M}_{23},\mathrm{M}_{3},\mathrm{M}_{13},\mathrm{M}_{1})(\ce{H3O+}\ce{OH-}). \label{eq:sigma_triprotic_11}
\end{align}
There are 12 permutations, the identity $e$ and the 11 permutations given by Equations \eqref{eq:sigma_triprotic_1}--\eqref{eq:sigma_triprotic_3} and \eqref{eq:sigma_triprotic_4}--\eqref{eq:sigma_triprotic_11}. These 12 permutations are divided in 6 tautomerizations and 6 acid-base permutations. The 6 tautomerizations are $e$, $\sigma_{\mathrm{d}}$, $\sigma_{\mathrm{d}}'$, $\sigma_{\mathrm{d}}''$, $c_3$, and $c_3^2$. The 6 acid-base permutations are $c_2$, $c_2'$, $c_2''$, $\mathrm{i}$, $s_6$, and $s_6'$. 

The graph automorphisms \eqref{eq:sigma_triprotic_4}-\eqref{eq:sigma_triprotic_11} can be obtained by composition of the generators $\sigma_\mathrm{d}$, $\sigma_\mathrm{d}'$, and $c_2$. A simple inspection gives
\begin{align}
    c_3&=\sigma_{\mathrm{d}}\sigma_{\mathrm{d}}',\\
    c_3^2&=\sigma_{\mathrm{d}}'\sigma_{\mathrm{d}},\\
    \sigma_{\mathrm{d}}''&=c_3\sigma_{\mathrm{d}} = c_3^2\sigma_{\mathrm{d}}',\\
    s_6&=c_2 \sigma_{\mathrm{d}}'=c_2'\sigma_{\mathrm{d}},\\
    s_6'&=c_2 \sigma_{\mathrm{d}}, \\
    c_2''&=\mathrm{i} \sigma_{\mathrm{d}}=s_6'\sigma_{\mathrm{d}}',\\
    \mathrm{i}&=c_2'\sigma_{\mathrm{d}}'.
\end{align}

\begin{figure}
    \centering
    \begin{tikzpicture}
        \Vertex[label=$e$, x=0, y=0,opacity=0.2,color=black]{e}        \Vertex[label=$\sigma_{\mathrm{d}}'$,x=0,y=-1.5,opacity=0.2,color=black]{sdp}        \Vertex[label=$\sigma_{\mathrm{d}}$,x=-1.3,y=0.75,opacity=0.2,color=black]{sd}
        \Vertex[label=$c_2$,x=1.08,y=0.625,opacity=0.2,color=black]{c2}
        \Vertex[label=$c_3$,x=-2.6,y=0,opacity=0.2,color=black]{c3}       \Vertex[label=$\sigma_{\mathrm{d}}''$,x=-2.6,y=-1.5,opacity=0.2,color=black]{sdpp}   \Vertex[label=$c_3^2$,x=-1.3,y=-2.25,opacity=0.2,color=black]{c32}
        \Vertex[label=$s_6$,x=-1.3,y=2,opacity=0.2,color=black]{s6}
        \Vertex[label=$c_2'$,x=-3.68,y=0.625,opacity=0.2,color=black]{c2p} 
        \Vertex[label=$\mathrm{i}$,x=-3.68,y=-2.125,opacity=0.2,color=black]{i} 
        \Vertex[label=$c_2''$,x=-1.3,y=-3.5,opacity=0.2,color=black]{c2pp} 
        \Vertex[label=$s_6'$,x=1.08,y=-2.125,opacity=0.2,color=black]{s6p} 
        
        \Edge[style={red},lw=2pt](e)(sd)
        \Edge[style={green},lw=2pt](sdp)(e)
        \Edge[style={blue},lw=2pt](e)(c2)
        \Edge[style={red},lw=2pt](c32)(sdp)
        \Edge[style={green},lw=2pt](sdpp)(c32)
        \Edge[style={red},lw=2pt](c3)(sdpp)
        \Edge[style={green},lw=2pt](sd)(c3)
        
        \Edge[style=blue,lw=2pt](sdp)(s6p)
        \Edge[style=green,lw=2pt](c2)(s6)
        \Edge[style=blue,lw=2pt](c3)(c2p)
        \Edge[style=red,lw=2pt](s6)(c2p)
        \Edge[style=blue,lw=2pt](sdpp)(i)
        \Edge[style=green,lw=2pt](c2p)(i)
        \Edge[style=blue,lw=2pt](c32)(c2pp)
        \Edge[style=red,lw=2pt](i)(c2pp)
        \Edge[style=blue,lw=2pt](sd)(s6)
        \Edge[style=red,lw=2pt](s6p)(c2)
        \Edge[style=green,lw=2pt](c2pp)(s6p)        
    \end{tikzpicture}
    \caption{Cayley graph for the 3-protic acid dissociation. Red, green and blue edges represent the action of the generators $\sigma_{\mathrm{d}}$, $\sigma_{\mathrm{d}}'$, and $c_2$, respectively.}
    \label{fig:cayley_graph_g3}
\end{figure}
In these compositions the product is taken from left to right. Other compositions, not shown here, also exist. A global view of the group $\mathrm{Aut}(G_3)$, with presentation given by Equation \eqref{eq:graph_aut_group_G3}, is given by the Cayley graph shown in Figure \ref{fig:cayley_graph_g3}. This graph features two hexagonal cycles, both generated by $\sigma_{\mathrm{d}}$ and $\sigma_{\mathrm{d}}'$, and connected by blue edges, $c_2$. The inner hexagon consists solely of tautomerizations, while the outer hexagon comprises acid-base permutations. The set of tautomerizations $H_T$, and the set $H_C=\{e,c_2\}$, both contain the identity $e$, hence $H_T$ and $H_C$ are subgroups of $\mathrm{Aut}(G_3)$. 
The subgroup $H_T$ is generated by $\sigma_{\mathrm{d}}$ and $\sigma_{\mathrm{d}}'$, and is isomorphic to the symmetric group $S_3$ \cite{Nash2005}. The subgroup $H_C$ is generated by $c_2$ and is isomorphic to the cyclic group $C_2$. The outer hexagon of the Cayley graph \ref{fig:cayley_graph_g3} is not a subgroup of $\mathrm{Aut}(G_3)$, it is the left coset $c_2H_T$ of $H_T$. It can be shown that $c_2 H_R = H_R c_2$, hence $H_R$ is a normal subgroup of $\mathrm{Aut}(G_3)$, $H_R\lhd\mathrm{Aut}(G_3)$. Since $C_2$ is also a normal subgroup of $\mathrm{Aut}(G_3)$, $\mathrm{Aut}(G_3)$ is the direct product of $S_3$ and $C_2$, $\mathrm{Aut}(G_3)=S_3\times C_2$. The 12 graph automorphisms, given by $e$ and the permutations of Equations \eqref{eq:sigma_triprotic_1}--\eqref{eq:sigma_triprotic_3} and \eqref{eq:sigma_triprotic_4}--\eqref{eq:sigma_triprotic_11}, are the elements of the group $\mathrm{Aut}(G_3) = S_3\times C_2$, which is isomorphic to $C_2\times S_3$, and to the abstract dihedral group $D_6$ and the antiprismatic 3D point group $D_{3\mathrm{d}}$ (Schoenflies' notation).   

\begin{figure}
    \centering
    \includegraphics[width=0.5\linewidth]{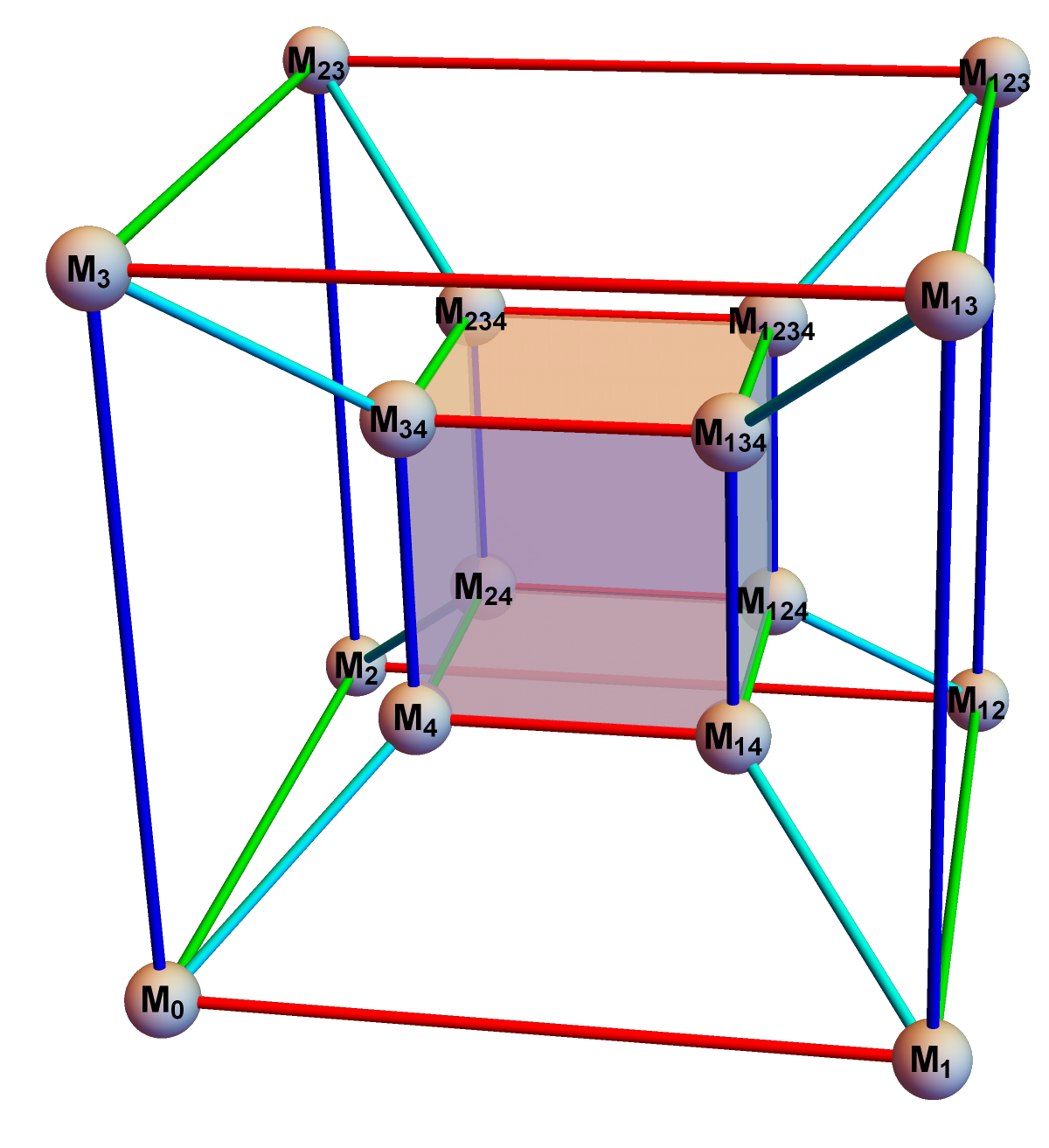}
    \caption{Graph $G_{k,4}=\left({V}_4,R_4\right)$ of the DMEs of the 4-protic acid. Red, green, blue and cyan edges represent the deprotonation/protonation of protons (1), (2), (3), and (4), respectively.}
    \label{fig:G_k4}
\end{figure}

The dissociation of the 4-protic acid is represented by the graph $G_4=\left(V_4,E_4\right)$, with ${V}_4=\{\mathrm{M}_\mu\,:\,\mu\in\mathcal{P}({\mathbb{S}}_4)\}$, and $E_4=R_4\cup T_4$. The set $R_4$ is the vertex set of DMSs related by micro-equilibrium constants, the set $T_4$ is the vertex set of pairs of DMSs related by tautomerization constants. The graph of the DMEs without tautomerizations, $G_{k,4}=\left({V}_4,R_4\right)$, is shown in Figure \ref{fig:G_k4}, which is a hypercube graph $Q_4$ with $2^4=16$ vertices and $4\times 2^{4-1}=32$ edges. The graph of the tautomerizations, $G_{\tau,4}=\left({V}_4,T_4\right)$, is shown in Figure \ref{fig:G_t4}, which is a disjoint graph made of two trivial graphs ($K_1$) for $\mathrm{M}_{\mathbb{S}_4}$ and $\mathrm{M}_\varnothing$, two tetrahedral graphs ($K_4$) for the DMSs of $\mathrm{Z}_1$ and $\mathrm{Z}_3$, and one octahedral graph ($K_{2,2,2}$) for the DMSs of $\mathrm{Z}_2$. 

\begin{figure}
    \centering
    \begin{tikzpicture}
        
        \Text[x=-2,y=2.25]{$\mathrm{Z}_0, K_1$}
        
        \Vertex[label=$\mathrm{M}_{\mathbb{S}_4}$, x=-2, y=1.5,opacity=0.2,color=black]{MS4}
        
        \Vertex[label=$\mathrm{M}_{124}$, x=-3, y=-1.5,opacity=0.2,color=black]{M124}
        \Vertex[label=$\mathrm{M}_{234}$, x=-3, y=0.5,opacity=0.2,color=black]{M234}
        \Vertex[label=$\mathrm{M}_{134}$, x=-1, y=0.5,opacity=0.2,color=black]{M134}
        \Vertex[label=$\mathrm{M}_{123}$, x=-1, y=-1.5,opacity=0.2,color=black]{M123}
        \Edge[style=gray](M124)(M234)
        \Edge[style=gray](M123)(M234)
        \Edge[style=gray](M234)(M134)
        \Edge[style=gray](M134)(M123)
        \Edge[style=gray](M123)(M124)
        \Edge[style=gray](M124)(M134)

        \Text[x=-2,y=-2]{$\mathrm{Z}_1, K_4$}
        
        \Vertex[label=$\mathrm{M}_{34}$, x=0, y=0,opacity=0.2,color=black]{M34}
        \Vertex[label=$\mathrm{M}_{13}$, x=1, y=2,opacity=0.2,color=black]{M13}
        \Vertex[label=$\mathrm{M}_{23}$, x=3, y=2,opacity=0.2,color=black]{M23}
        \Vertex[label=$\mathrm{M}_{12}$, x=4, y=0,opacity=0.2,color=black]{M12}
        \Vertex[label=$\mathrm{M}_{24}$, x=3, y=-2,opacity=0.2,color=black]{M24}
        \Vertex[label=$\mathrm{M}_{14}$, x=1, y=-2,opacity=0.2,color=black]{M14}
        \Edge[style=gray](M34)(M13)
        \Edge[style=gray](M13)(M23)
        \Edge[style=gray](M23)(M12)
        \Edge[style=gray](M12)(M24)
        \Edge[style=gray](M24)(M14)
        \Edge[style=gray](M14)(M34)
        \Edge[style=gray](M34)(M23)
        \Edge[style=gray](M34)(M24)
        \Edge[style=gray](M13)(M14)
        \Edge[style=gray](M13)(M12)
        \Edge[style=gray](M23)(M24)
        \Edge[style=gray](M12)(M14)

        \Text[x=2,y=0]{$\mathrm{Z}_2, K_{2,2,2}$}

        \Vertex[label=$\mathrm{M}_{2}$, x=5, y=0.5,opacity=0.2,color=black]{M2}
        \Vertex[label=$\mathrm{M}_{4}$, x=5, y=-1.5,opacity=0.2,color=black]{M4}
        \Vertex[label=$\mathrm{M}_{3}$, x=7, y=-1.5,opacity=0.2,color=black]{M3}
        \Vertex[label=$\mathrm{M}_{1}$, x=7, y=0.5,opacity=0.2,color=black]{M1}
        \Edge[style=gray](M1)(M2)
        \Edge[style=gray](M2)(M4)
        \Edge[style=gray](M3)(M4)
        \Edge[style=gray](M3)(M1)
        \Edge[style=gray](M1)(M4)
        \Edge[style=gray](M3)(M2)
        \Text[x=6,y=-2]{$\mathrm{Z}_3, K_4$}

        \Vertex[label=$\mathrm{M}_{\varnothing}$,x=6, y=1.5,opacity=0.2,color=black]{M0}

        \Text[x=6,y=2.25]{$\mathrm{Z}_4, K_1$}
    \end{tikzpicture}
    \caption{Graph $G_{\tau,4}$ of the tautomerizations of the 4-protic acid. This is a disjoint graph made of four subgraphs. $\mathrm{M}_{\mathbb{S}_4}$ and $\mathrm{M}_0$ are modeled by trivial graphs, the tautomers of $\mathrm{Z}_1$ and $\mathrm{Z}_3$ are modeled by tetrahedral graphs, finally the tautomerizations of $\mathrm{Z}_2$ are modeled by the octahedral graph.}
    \label{fig:G_t4}
\end{figure}
The graph automorphism group $\mathrm{Aut}(G_4)$ is obtained with the aid of Wolfram Mathematica 12 \cite{Mathematica12} through
\begin{equation}\label{eq:Aut_G_4}
    \mathrm{Aut}(G_4)=\texttt{GraphAutomorphismGroup[}\includegraphics[scale=0.4,valign=m]{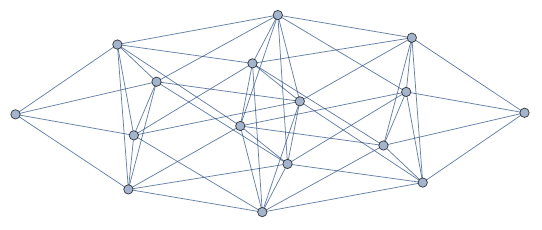}\texttt{]},
\end{equation}
which gives a set of generators for $\mathrm{Aut}(G_4)$:
\begin{align}
\sigma_1=&\left(\mathrm{M}_{124},\mathrm{M}_{134}\right)\left(\mathrm{M}_{12},\mathrm{M}_{13}\right)\left(\mathrm{M}_{24},\mathrm{M}_{34}\right)\left(\mathrm{M}_{2},\mathrm{M}_{3}\right),\\
\sigma_2=&\left(\mathrm{M}_{134},\mathrm{M}_{234}\right)\left(\mathrm{M}_{13},\mathrm{M}_{23}\right)\left(\mathrm{M}_{14},\mathrm{M}_{24}\right)\left(\mathrm{M}_{1},\mathrm{M}_{2}\right),\\
\sigma_3=&\left(\mathrm{M}_{123},\mathrm{M}_{124}\right)\left(\mathrm{M}_{13},\mathrm{M}_{14}\right)\left(\mathrm{M}_{23},\mathrm{M}_{24}\right)\left(\mathrm{M}_{3},\mathrm{M}_{4}\right),\\
\sigma_4=&\left(\mathrm{M}_{\mathbb{S}_4},\mathrm{M}_{\varnothing}\right)\left(\mathrm{M}_{123},\mathrm{M}_{1}\right)\left(\mathrm{M}_{124},\mathrm{M}_{2}\right)\left(\mathrm{M}_{134},\mathrm{M}_{3}\right)\\&\left(\mathrm{M}_{234},\mathrm{M}_{4}\right)\left(\mathrm{M}_{14},\mathrm{M}_{23}\right)\nonumber(\ce{H3O+}\ce{OH-}).
\end{align}

These permutations and the identity $e$ generate a total of 48 group elements. The multiplication table (or Cayley table) of a group encodes all the information about the group's operation. Specifically, it shows how every pair of elements in the group combines under the group operation. Two groups are isomorphic if they have identical multiplication tables or if there exists a set of permutations of the elements of one group that transforms its multiplication table into a table identical to that of the other group \cite{carter2009}. The multiplication table of $\mathrm{Aut}(G_4)$, $\mathcal{M}_4$, is obtained using  Mathematica 12 with the command:
\begin{equation}
\mathcal{M}_4=\texttt{GroupMultiplicationTable[}\mathrm{Aut}(G_4)\texttt{]}.
\end{equation}

The multiplication table of the direct product group $C_2\times S_4$ is obtained with the command: 
\begin{equation}
\begin{split}
    \mathcal{M}_{C_2\times S_4}=&\texttt{FiniteGroupData[} \\
    &\,\texttt{\{"DirectProduct",\{\{"CyclicGroup",\,2\},\{"SymmetricGroup",\,4\}\}\},}\\ &\,\texttt{"MultiplicationTable"]}.    
\end{split}
\end{equation}

\begin{figure}
    \centering
    \includegraphics[scale=0.6]{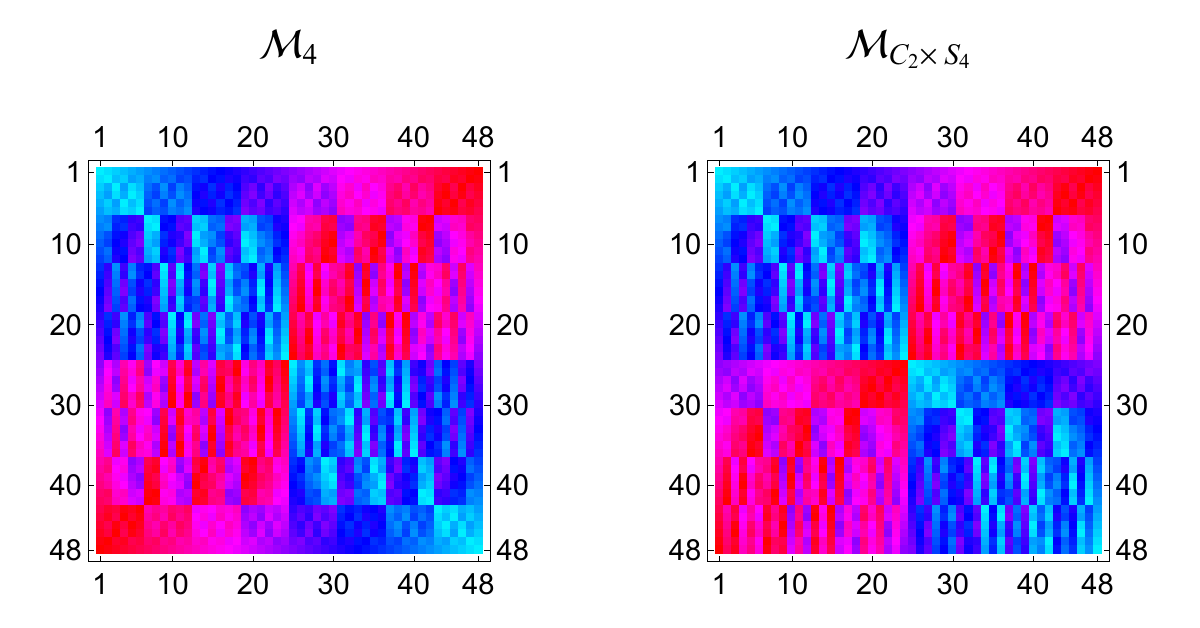}
    \caption{\texttt{MatrixPlot} of the multiplication tables of $\textrm{Aut}(G_4)$, $\mathcal{M}_4$, and of the direct product $C_2 \times S_4$, $\mathcal{M}_{C_2 \times S_4}$. }
    \label{fig:mt_M_4_vs_M_C2xS4}
\end{figure}

The elements of these matrices are integer numbers ranging from $1$ to $48$, each representing an element of the respective groups. Figure \ref{fig:mt_M_4_vs_M_C2xS4} displays the multiplication tables $\mathcal{M}_4$ and $\mathcal{M}_{C_2 \times S_4}$, visualized using Mathematica's $\texttt{MatrixPlot}$ function with a hue-based color scheme. These multiplication tables each consist of four blocks. All the blocks in $\mathcal{M}_{C_2\times S_4}$ and the two top blocks in $\mathcal{M}_4$ are isomorphic to the Cayley table of the group $S_4$, $\mathcal{M}_{S_4}$. The bottom blocks of $\mathcal{M}_4$ can be transformed into the corresponding blocks of $\mathcal{M}_{C_2\times S_4}$ by reversing the rows and columns within each block. More precisely, there exists 
a permutation matrix $\mathsf{P}$ that transforms the bottom blocks of $\mathcal{M}_4$ into matrices isomorphic to $\mathcal{M}_{S_4}$ after a suitable renaming of the elements. This matrix $\mathsf{P}$ is given by the $2\times 2$ block matrix
\begin{equation}
    \mathsf{P}=\begin{pmatrix}
        \mathsf{I}_{24} & \mathsf{0}_{24} \\
        \mathsf{0}_{24} & \mathsf{J}_{24}
    \end{pmatrix}.
\end{equation}
In this matrix, $\mathsf{I}_{24}$ and  $\mathsf{0}_{24}$ are the $24\times 24$ identity and zero matrices, respectively, and $\mathsf{J}_{24}$ is the $24\times 24$ exchange matrix given by
\begin{equation}
    \mathrm{J}_{24}=\delta_{i,25-j}, \hspace{1cm} 1 \le i,j \le 24,
\end{equation}

with $\delta_{i,j}$ as the Kronecker delta which is 1 if $i=j$ and zero otherwise. The matrix $\mathsf{P}$ transforms $\mathcal{M}_4$ into $\mathcal{M}_4'$ using the equation
\begin{equation}
    \mathcal{M}_4'=\mathsf{P}\mathcal{M}_4\mathsf{P}^\intercal.
\end{equation}
Next, by applying the relabeling rule
\begin{equation}
    \mathcal{R}=\left\{48\to 25, 47 \to 26, \dots,25\to 48\right\},
\end{equation}
the matrix $\mathcal{M}_4'$ is converted into $\mathcal{M}_4''$ which is isomorphic to $\mathcal{M}_{C_2\times S_4}$. Consequently $\mathrm{Aut}(G_4)\cong C_2\times S_4$, which is isomorphic to the octahedral group $O_h$. 

\begin{figure}
    \centering
    \includegraphics[scale=0.6]{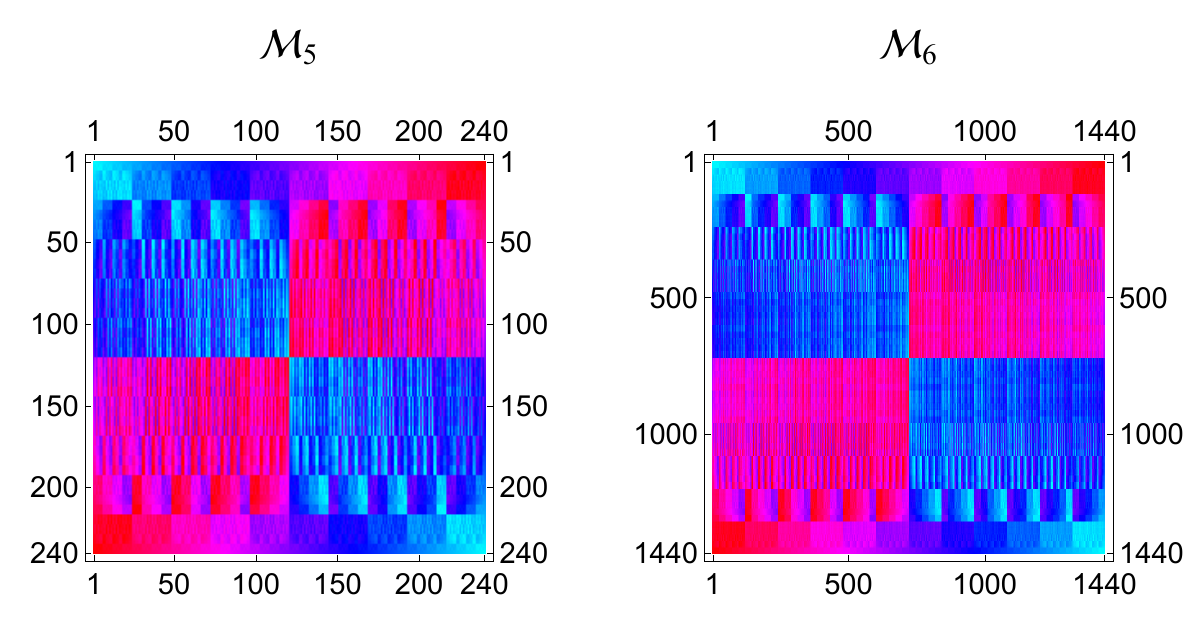}
    \caption{\texttt{MatrixPlot} of the Cayley tables of the groups $\mathrm{Aut}(G_5)$ and $\mathrm{Aut}(G_6)$.}
    \label{fig:mt_g5_g6}
\end{figure}
The Cayley tables for $\mathrm{Aut}(G_5)$ and $\mathrm{Aut}(G_6)$, denoted $\mathcal{M}_5$ and $\mathcal{M}_6$ respectively, are shown in the Figure \ref{fig:mt_g5_g6}. These tables exhibit a similar pattern to the Cayley table $\mathcal{M}_4$ represented in Figure \ref{fig:mt_M_4_vs_M_C2xS4}.  The Cayley tables for the symmetric groups $S_5$ and $S_6$ can be generated in Mathematica 12 using the command:
\begin{equation}
\mathcal{M}_{S_n}=\texttt{GroupMultiplicationTable[SymmetricGroup[n]]},
\end{equation}
with $\texttt{n}=5,6$. The top blocks of $\mathcal{M}_5$ and $\mathcal{M}_6$ are isomorphic to $\mathcal{M}_{S_5}$ and $\mathcal{M}_{S_6}$ respectively. There are specific matrices and renaming rules that transform the bottom blocks of  $\mathcal{M}_5$ and $\mathcal{M}_6$ into matrices isomorphic to $\mathcal{M}_{S_5}$ and $\mathcal{M}_{S_6}$ respectively. These matrices are given by 
\begin{equation}
    \mathsf{P}_n=\begin{pmatrix}
        \mathsf{I}_{n!} & \mathsf{0}_{n!} \\
        \mathsf{0}_{n!} & \mathsf{J}_{n!}
    \end{pmatrix},
\end{equation}
with $\mathsf{I}_{n!}$ and $\mathsf{0}_{n!}$ as the $n!$-dimensional identity and zero matrices, and $\mathsf{J}_{n!}$ as the  $n!$-dimensional exchange matrix explicitly defined by
\begin{equation}
     \mathsf{J}_{n!}=\delta_{i,1+n!-j},\hspace{1cm}1\le i,j \le n!.
\end{equation}
The renaming rules are given by 
\begin{equation}
    \mathcal{R}_n=\left\{ 2n!\to n!+1, 2n!-1\to  n!+2, \dots,n!+1 \to 2n! \right\}.
\end{equation}
The use of these matrices and renaming rules on the multiplication tables $\mathcal{M}_5$ and $\mathcal{M}_6$ gives matrices $\mathcal{M}_5''$ and $\mathcal{M}_6''$ that are isomorphic to $\mathcal{M}_{S_5}$ and $\mathcal{M}_{S_6}$. respectively.

It has been shown that for $N=1,2,\dots,6$, the microdissociation of an $N$-protic acid has a graph $G_N$ with graph automorphism group 
\begin{equation}
    \mathrm{Aut}(G_N)\cong C_2\times S_N.
\end{equation}
The observed behavior is attributed to the interplay of two distinct equilibrium types: microdissociation and tautomerization. Microdissociation equilibria correlate with the $C_2$ component, while tautomerization equilibria are linked to the $S_N$ component of the automorphism group. 

Based on the underlying physics and chemistry of the microdissociation, we hypothesize that the graph representing this process has an automorphism group isomorphic to the direct product $C_2\times S_N$. 

\section{Concluding remarks}

The description of the microdissociation of an $N$-protic acid in terms of set theory is utilized to derive mathematical relations between equilibrium and micro-equilibrium dissociation constants. These mathematical relations are more convenient to use compared to the formulas based on indexation provided by Hill \cite{Hill1943}. The advantages of our formalism are demonstrated by deriving equations that relate the equilibrium and micro-equilibrium constants of diprotic and triprotic acids. 

Graph theory has been employed to represent and classify the microdissociation equilibrium of polyprotic acids as graphs, denoted as $G_N$, with the dissociation micro-states as vertices and the pairs of vertices connected by microdissociation constants as edges. Tautomerizations and acid-base reactions are treated as permutations on the vertex set of the polyprotic acid graph. It is shown that these permutation are graph automorphisms, and the composition of two of these permutations is also a graph automorphism. The set of automorphisms, endowed with composition, forms the automorphism group of the graph $G_N$. The generators of these groups were completely identified for monoprotic to 4-protic acids. In the case of monoprotic acids the graph automorphism group is given by the cyclic group $C_2\cong S_2$. The analysis of the dissociation of a diprotic acid reveals the direct product $C_2\times C_2$. An analysis of the Cayley graph of $\mathrm{Aut}(G_3)$ allowed the identification of $C_2$ and $S_3$ as normal subgroups, leading to the conclusion that $\mathrm{Aut}(G_3)$ is isomorphic to the direct product $C_2\times S_3$. The Cayley tables of $\mathrm{Aut}(G_4)$ and $C_2\times S_4$, $\mathcal{M}_4$ and $\mathcal{M}_{C_2\times S_4}$ respectively, were compared. These tables exhibit a $2\times 2$ block structure, with the top blocks isomorphic to the multiplication table of the symmetric group $\mathcal{M}_{S_4}$. By permuting pairs of elements of $\mathcal{M}_4$ and renaming some elements, a multiplication table isomorphic to $\mathcal{M}_{S_4}$ was obtained, confirming that $\mathrm{Aut}(G_4)$ is isomorphic to the direct product $C_2\times S_4$. Furthermore, the multiplication tables of the graph automorphism groups of the 5-protic and 6-protic acids were compared to those of the symmetric groups $S_5$ and $S_6$. The same block structure observed in $\mathcal{M}_4$ was found in these comparisons. Computational verification confirmed that each of the four blocks of these multiplication tables is isomorphic to either $\mathcal{M}_{S_5}$ or $\mathcal{M}_{S_6}$, indicating that $\mathrm{Aut}(G_N)\cong C_2\times S_N$ for $N=1,2,\dots,6$.

The computational proof for the 6-protic acid is particularly significant due the unique nature of the automorphism group of $S_6$ \cite{Janusz1982,Fournelle1993}. The automorphism group of $S_6$ is isomorphic to the semidirect product $S_6 \rtimes C_2$, which includes an additional outer automorphism beyond the inner automorphisms derived from $S_6$'s elements. It is important to note that the multiplication table of $S_6$ captures only the inner automorphisms, reflecting how elements relate through conjugation, and does not account for the outer automorphism present in $\mathrm{Aut}(S_6)$. Since the 4 blocks of the multiplication table $\mathcal{M}_6$ are isomorphic to the multiplication table $\mathcal{M}_{S_6}$, the outer automorphism is irrelevant for our purposes, as the table primary highlights the inner group structure, not additional symmetries introduced by outer automorphisms. 

The formalism and results presented in this paper contribute to advance our comprehension of acid-base equilibria and provide a foundation for future investigations into more complex chemical systems, such as biochemical and macromolecular processes.

\section*{Author Contributions}

CAA led the conceptualization, formal analysis, funding acquisition, investigation, model development, and writing of the original draft. NS and JL handled data curation, investigation, computer program design, and figure creation, and they also assisted with text revision.

\section*{Conflict of Interest}

No potential conflict of interest was reported by the authors.

\section{Acknowledgments}

This work has been partially funded by OMICAS Program: In-silico Multiscale Optimization of Sustainable Agricultural Crops (Infrastructure and validation in Rice and Sugar Cane) sponsored within the Scientific Colombia Ecosystem, made up by the World Bank, Ministry of Science, Technology, and Innovation (Minciencias), Icetex, Ministry of Education and Ministry of Industry and Tourism, Project ID: FP44842-217-2018. Partial funding was also provided by the internal research grants of Universidad Icesi.

The authors express their sincere gratitude to Professor Mathew Macauley of Clemson University for his invaluable insights and generous correspondence regarding group theory and graphical representation of groups.  

\section*{Data Availability Statement}

The results of this work are mainly analytical. Codes to produce the figures and Wolfram Mathematica Notebooks are available at \url{https://github.com/caarango/TizZ-codes-} or can be obtained by request from the corresponding author.

\bibliography{references}

\end{document}